\documentclass[journal]{IEEEtran}
\usepackage{citesort,amssymb,amsmath,amsthm,graphicx}
\usepackage[T1]{fontenc}

\theoremstyle{definition}
\newtheorem{PCO-ST}{Theorem}
\newtheorem{PSO-ST}[PCO-ST]{Theorem}
\newtheorem{PCO-INT}{Lemma}
\newtheorem{PCO-AN}[PCO-ST]{Theorem}
\newtheorem{PCO-LOW-AN}{Corollary}
\newtheorem{PSO-AN}[PCO-ST]{Theorem}
\newtheorem{STC-ST}{Proposition}
\newtheorem{PHI-OPT-ST}[PCO-LOW-AN]{Corollary}

\begin{document}
\title{Enhancing Secrecy with Multi-Antenna Transmission in Wireless Ad Hoc Networks}
\author{
Xi~Zhang, \IEEEmembership{Student Member, IEEE,} Xiangyun~Zhou, \IEEEmembership{Member, IEEE,}\\and Matthew~R.~McKay, \IEEEmembership{Senior Member, IEEE}
\thanks{Copyright~(c) 2013 IEEE. Personal use of this material is permitted. However, permission to use this material for any other purposes must be obtained from the IEEE by sending a request to pubs-permissions@ieee.org.}
\thanks{X.~Zhang and M.~R.~McKay are with the Department of Electronic and Computer Engineering, the Hong Kong University of Science and Technology, Hong Kong~(e-mails: xizhangx@ust.hk, eemckay@ust.hk).}
\thanks{X.~Zhou is with the Research School of Engineering, the Australian National University, Australia~(e-mail: xiangyun.zhou@anu.edu.au).}
\thanks{This paper was presented in part at the IEEE International Workshop on Signal Processing Advances for Wireless Communications, Darmstadt, Germany, June 16--19, 2013. The work of X.~Zhang and M.~R.~McKay was supported by the Hong Kong Research Grants Council~(Grant No. 616312). The work of X.~Zhou was supported by the Australian Research Council's Discovery Projects funding scheme~(Project No. DP110102548).}
}
\maketitle

\begin{abstract}
We study physical-layer security in wireless ad hoc networks and investigate two types of multi-antenna transmission schemes for providing secrecy enhancements. To establish secure transmission against malicious eavesdroppers, we consider the generation of artificial noise with either sectoring or beamforming. For both approaches, we provide a statistical characterization and tradeoff analysis of the outage performance of the legitimate communication and the eavesdropping links. We then investigate the networkwide secrecy throughput performance of both schemes in terms of the secrecy transmission capacity, and study the optimal power allocation between the information signal and the artificial noise. Our analysis indicates that, under transmit power optimization, the beamforming scheme outperforms the sectoring scheme, except for the case where the number of transmit antennas are sufficiently large. Our study also reveals some interesting differences between the optimal power allocation for the sectoring and beamforming schemes.
\end{abstract}


\section{Introduction}
\IEEEPARstart{I}{nformation} security is a prime concern in emerging wireless networks. Traditional security mechanisms, typically involving cryptographic algorithms, implicitly assume that any potential eavesdroppers have limited computational abilities. Rapid development of computing hardware, however, has meant that there is an ever-increasing susceptibility of such methods to attack. To address this problem and further strengthen existing wireless security technologies, physical-layer security mechanisms~\cite{Wyner1975,Csiszar1978,Bloch2011} have recently attracted significant attention. Such mechanisms work by exploiting the physical properties of the wireless medium in order to provide an additional layer of robustness which is ``information-theoretically secure''. In the past decade, physical-layer security techniques have been proposed in many different communication scenarios; e.g., multi-input multi-output, relay/jammer-assisted, cognitive-radio-enabled, and so on~(see e.g.,~\cite{Oggier2008,Wang2010,Huang2011,Zhang2012,Mukherjee2013,Luo2013} and the references therein). Very recently, physical-layer security techniques have also been introduced into large-scale decentralized ad hoc networks, in order to provide enhanced secrecy performance~\cite{Liang2009,Zhou2011A,Pinto2012A,Zhou2011B,Lee2012,Koyluoglu2010,Vasudevan2010,Zhou2012,Shi2012}.

When characterizing the networkwide secrecy throughput of large-scale networks, the interference due to concurrent transmissions plays an important role. On the one hand, from the point of view of the legitimate receivers, interference is a nuisance which leads to unwanted throughput loss; whilst, on the other hand, to the eavesdroppers, interference makes it more difficult to intercept transmissions~(thereby providing enhanced security). In decentralized networks, it is reasonable to assume that the legitimate receivers do not have sophisticated multi-user decoding capabilities, and thereby treat interference simply as noise. For the eavesdroppers, however, one should typically not make such strong assumptions, since the capabilities of eavesdroppers are often unknown. This has led researchers to design for a worst-case scenario~(see e.g.,~\cite{Koyluoglu2010,Vasudevan2010,Zhou2012}), whereby the eavesdroppers are assumed to be capable of performing multi-user decoding~(e.g., successive interference cancellation), allowing the interference created by concurrent transmission of information signals to be potentially resolved.

In order to confuse eavesdroppers with multi-user decodability in ad hoc networks, methods have been introduced in~\cite{Vasudevan2010,Zhou2012} based on generating artificial noise~(see~\cite{Goel2008,Swindlehurst2009}) or cooperative jamming~(see~\cite{Tekin2008,He2010}). In both cases, the aim is to create non-resolvable interference at the eavesdroppers. Specifically, in~\cite{Vasudevan2010}, the legitimate users which are far away from the intended receiver were selected to emit artificial noise; while in~\cite{Zhou2012}, a certain percentage of legitimate users were randomly chosen to radiate jamming signals. Note that when the legitimate nodes have only a single antenna, as in~\cite{Vasudevan2010,Zhou2012}, some legitimate users must suspend their own message transmission in order to deliver artificial noise or jamming signals. In~\cite{Shi2012}, the authors considered the case where there are many multi-antenna helping jammers, generating artificial interference but zero-forcing to nearby legitimate receivers. The injected jamming signals can also help to deal with eavesdroppers with multi-user decodability. However, in some cases, such a large number of helping jammers may not be obtainable and when this happens, the designed transmission scheme might be vulnerable.

\subsection{Our Approach and Contribution}
In this paper, we consider two artificial-noise-aided multi-antenna transmission strategies, based on antenna sectoring or beamforming, for providing secrecy in wireless ad hoc networks. Knowing the direction of the intended receiver, with antenna sectoring, an information signal is transmitted towards the intended receiver, while artificial noise is simultaneously radiated in other sectors. With the channel state information~(CSI) of the intended receiver, through beamforming, artificial noise is injected into the null-space of the intended channel, leaving the intended receiver unaffected. Note that these two transmission schemes do \emph{not} require any instantaneous channel knowledge of the eavesdroppers. With the proposed transmission schemes, non-resolvable interference is created at the potential eavesdroppers by the introduced artificial noise, allowing us to guarantee a target secrecy throughput, which appears difficult without using artificial noise. Moreover, as a major advantage of these schemes, no legitimate users must stop their own message transmission, which is in contrast to the single-antenna transmission approaches~\cite{Vasudevan2010,Zhou2012}. In particular, since the artificial-noise generated interference is created by the legitimate transmitters themselves, the system can work in the absence of helping jammers~\cite{Shi2012}.

Previous studies in~\cite{Goel2008,Tekin2008,Swindlehurst2009,He2010,Zhou2010,Liao2011,Zhang2011,Lin2011,Li2011,Romero-Zurita2012,Gerbracht2012,Zhang2013A} have clearly shown that smartly injecting artificial interference can achieve secrecy enhancements in point-to-point scenarios. In such cases, the artificial noise generated from multi-antenna beamforming results in interference in some subspaces but no interference in others. Contrarily, in ad hoc networks where transmitters generate artificial noise in a decentralized manner, artificial interference is created collectively in all subspaces over the entire network. In short, the artificial-noise-aided beamforming and sectoring schemes considered in this paper have a \emph{cooperative jamming} effect in ad hoc networks, which ensures that every eavesdropper is subjected to jamming. With these cooperative jamming effects, it is unclear how well the proposed artificial-noise-aided multi-antenna transmission schemes can work in large-scale networks, as a means of jointly providing high communication performance and security. Our main objective is to address this key question.

For both the sectoring and beamforming schemes, we characterize the performance of the legitimate link and the eavesdropping link, by deriving new closed-form expressions for the connection and secrecy outage probabilities respectively. We then illustrate the tradeoff between the connection and secrecy outage performance. Under constraints on the connection and secrecy outage probabilities, we quantify the achievable secrecy throughput performance of both schemes in terms of the secrecy transmission capacity. Finally, we investigate the optimal power allocation between the information signal and the artificial noise which maximizes the secrecy transmission capacity, and compare the corresponding maximum secrecy transmission capacity of both schemes. Our analytical and numerical results suggest that the considered multi-antenna transmission schemes can achieve significant secrecy enhancements in wireless ad hoc networks.

We observe that by adding extra transmit antennas, the optimal ratio of power allocated to the information signal that maximizes the secrecy transmission capacity of the beamforming scheme converges to a certain fraction which is strictly less than one, while that for the sectoring scheme keeps increasing towards one. We show that as the number of transmit antennas grows large, for both the sectoring and beamforming schemes, the optimized secrecy transmission capacity increases logarithmically. Our analysis also indicates that, in terms of the optimized secrecy transmission capacity, the beamforming scheme outperforms the sectoring scheme for a wide range of antenna numbers, at the expense of requiring additional channel knowledge. Nevertheless, as the number of transmit antennas becomes sufficiently large, quite interestingly, the sectoring scheme can achieve a better throughput performance than the beamforming scheme. The performance crossover happens at a larger antenna number if the secrecy outage constraint becomes more stringent.

The rest of the paper is organised as follows. In Section~\ref{sec:system-model}, we introduce the system model and illustrate the proposed transmission schemes. In Section~\ref{sec:outage-sectoring}, we characterize the outage performance of the sectoring scheme. In Section~\ref{sec:outage-beafroming}, we study the outage performance of the beamforming scheme. In Section~\ref{sec:throughput}, we analyze and compare the secrecy throughput performance of the sectoring and beamforming schemes. Finally, in Section~\ref{sec:conclusion}, we conclude this paper.

\section{System Model}\label{sec:system-model}
The legitimate transmitters and malicious eavesdroppers are modeled by two independent homogeneous Poisson point processes~(PPPs) on a two dimensional plane~$\mathbb{R}^2$ with densities~$\lambda_l$ and~$\lambda_e$, respectively. The location sets of the legitimate transmitters and eavesdroppers are denoted by~$\Phi_L$ and~$\Phi_E$, respectively. Following the widely-used bipolar network model~\cite{Haenggi2009A,Blaszczyszyn2010}, we assume that every transmitter has an intended receiver at distance~$r$ in a random direction\footnote{This bipolar network model is suitable for modeling the case where the distance from the transmitter to the intended receiver is relatively small, compared with the distances between the transmitters, and it is widely used in the literature~(see e.g.,~\cite{Zhou2011B,Zhou2012,Shi2012,Haenggi2009A,Blaszczyszyn2010}). Generalization can be made by setting the distance~$r$ as a random variable following a certain distribution, and then averaging the network performance over the distribution of~$r$, as mentioned in~\cite{Haenggi2009A}.}. Each transmitter is equipped with~$N$ transmit antennas, while each receiver~(both legitimate and malicious) has a single receive antenna. In addition to an exponential path loss with parameter~$\alpha>2$, the wireless channels are assumed to be experiencing independent Rayleigh fading. Since we are studying large-scale networks with uncoordinated concurrent transmissions, the aggregate interference will be dominant, and the local thermal noise is usually negligible. As done in~\cite{Zhou2011B,Zhou2012,Lee2012}, we omit the thermal noise and use the signal-to-interference ratio~(SIR) as the main performance metric.

The total transmit power at each transmitter is denoted by~$P$. Define~$\phi$ as the ratio of the power of the information signal to the total transmit power. Thus, the power allocated to the information signal is~$P_I=P\phi$, while the power allocated to the artificial noise is~$P_A=P\left(1-\phi\right)$. In the following, we consider the use of artificial noise in the form of either sectoring or beamforming, for providing secrecy against the malicious eavesdroppers.

\subsection{Sectoring with Artificial Noise}
With~$N$ directional antennas, each transmitter can send independent signals in~$N$ disjoint sectors, each of these covering~$\frac{2\pi}{N}$ radians with an antenna gain~$G_N$. The antenna gain usually increases as the spread angle decreases~(i.e., increasing~$N$). As done in~\cite{Pinto2012A}, we assume that the sidelobes are suppressed sufficiently and thus can be omitted in later analysis\footnote{We point out that the analytical results can be generalized to incorporate the sidelobe leakage signals, by accounting for the eavesdroppers outside the main lobe of the intended sector, and carefully evaluating the aggregate artificial noise received by the eavesdroppers. However, the resulting analytical expressions become much more complicated, whilst few new insights can be extracted. In fact, it can be shown that, as long as the leakage signals in the sidelobes are not particularly strong, the resulting performance degradations are insignificant. For simplicity, we focus on the case of perfectly sectorized antennas with negligible sidelobes.}. We then consider the following scheme to combine sectoring with artificial noise generation: Each transmitter sends an information signal in the sector containing its intended receiver, while simultaneously emitting artificial noise in all other sectors, creating non-resolvable interference to the malicious eavesdroppers. Note that the transmitter needs to know the direction of the intended receiver and this information can be accurately obtained through the discovery mechanisms, such as the ``informed discovery'' mechanism in~\cite{Ramanathan2005}, where the feedback from the intended receiver is exploited for a high accuracy. We further assume that the time needed for acquiring and feeding back this directional information is negligible. Note that this sectoring scheme does \emph{not} require the CSI at the transmitter. With the antenna gain, in the intended sector, the information signal is transmitted with power~$G_NP_I$, while in each of the other~$N-1$ sectors, artificial noise is radiated with power~$\frac{G_NP_A}{N-1}$.

\subsection{Beamforming with Artificial Noise}
With~$N$ omnidirectional antennas, the transmitters can perform artificial-noise-aided beamforming~\cite{Goel2008,Swindlehurst2009}. To do this, the instantaneous CSI of the intended receiver is needed at the associated transmitter. This information can be obtained via pilot training and we assume that the time needed for the training phase is negligible. With the CSI, the transmitter performs maximal ratio transmission towards the intended receiver with power~$P_I$. Meanwhile, to confuse the malicious eavesdroppers, artificial noise with total power~$P_A$ is uniformly injected into the null space of the intended channel. To be specific, denoting the intended channel by~${\bf h}$, an orthonormal basis is generated at the transmitter as~$\left[\frac{{\bf h}}{\|{\bf h}\|},{\bf W}\right]$, where~${\bf W}$ is a~$N\times\left(N-1\right)$ matrix, the columns of which are mutually orthogonal while also being orthogonal to~$\frac{{\bf h}}{\|{\bf h}\|}$. The message vector to be transmitted will then have the following form:
\begin{align}
{\bf x}=\frac{{\bf h}}{\|{\bf h}\|}u+{\bf W}{\bf v}\label{eq:an-an}
\end{align}
where~$u$ is the information signal, assumed to be complex Gaussian distributed with variance~$P_I$;~${\bf v}$ is the artificial noise vector, the entries of which are complex Gaussian distributed with zero mean and variance~$\sigma_v^2=\frac{P_A}{N-1}$.

\subsection{Wiretap Coding and Outage Definition}\label{sec:code-outage}
Here we explain the coding scheme and the outage definitions. Before transmission, the data is encoded using the wiretap code~\cite{Wyner1975}. The codeword rate and the secret message rate are denoted by~$R_b$ and~$R_s$, respectively. The codeword rate~$R_b$ is the actual transmission rate of the codewords, while the secrecy rate~$R_s$ is the rate of the embedded message. The rate redundancy~$R_e:=R_b-R_s$ is intentionally added, in order to provide secrecy against malicious eavesdropping. More discussions on code construction can be found in~\cite{Thangaraj2007}. If the channel from the transmitter to its intended receiver cannot support the codeword rate~$R_b$, the receiver may not be able to decode the transmitted codeword correctly. We consider this as a \emph{connection outage} event.

There are possibly several eavesdroppers trying to intercept the same transmitter, while the exact number of them is unknown. To minimize the assumption on the eavesdroppers' behavior and design for a worst case, we consider the scenario where all eavesdroppers are trying to decode the message from the transmitter under consideration. Therefore, if the channel from the transmitter to any of the eavesdroppers can support a data rate larger than the embedded rate redundancy~$R_{e}$, this transmission fails to achieve perfect secrecy and a \emph{secrecy outage} is deemed to occur~\cite{Zhou2011C}. As mentioned in~\cite{Lee2012}, this kind of secrecy outage formulation provides a ``strong'' secrecy performance.

Note that we assume the legitimate receivers do not apply multi-user decoding techniques. Thus, the interference at the legitimate receivers consists of the information signals and the artificial noise. On the other hand, we assume that the eavesdroppers are capable of multi-user decoding, i.e., resolving concurrent transmissions. In order to design the network parameters to achieve the maximum level of secrecy, as done in~\cite{Vasudevan2010,Zhou2012,Koyluoglu2010}, we consider a worst-case assumption to overestimate the eavesdroppers' multi-user decodability: For the signal reception at any eavesdropper, only the artificial noise constitutes the interference, whereas the received information signals are resolvable and hence are not part of the interference.

\section{Outage Performance of Sectoring Scheme}\label{sec:outage-sectoring}
In this section, we study the outage performance of the sectoring scheme. We start by deriving the outage probability of the intended links. Then, we characterize the possibility that the transmitted message is not secure against the malicious eavesdroppers.

\subsection{Connection Outage Probability}
Here, we derive the connection outage probability~$p_{{\rm co}}$. A threshold SIR value for connection outage is defined as~$\beta_{b}$. We focus on a typical transmitter-receiver pair and place the receiver at the origin of the coordinate system. From Slivnyak's theorem~\cite{Haenggi2009B}, the distribution of all other nodes' locations will not be affected; thus, the obtained statistics can reflect the system performance accurately.

For the typical receiver at the origin, the interfering nodes can be classified into two classes: 1) interferers transmitting information signals towards the typical receiver; 2) interferers sending artificial noise towards the typical receiver. With such a classification, by~\cite{Haenggi2009B}, the transmitters in~$\Phi_L$ can be divided into two independent homogeneous PPPs, which are denoted as~$\Phi_I$ and~$\Phi_A$ with densities~$\frac{1}{N}\lambda_l$ and~$\frac{N-1}{N}\lambda_l$, respectively.

Then, the aggregate interference resulting from the transmitters in~$\Phi_I$ and~$\Phi_A$ is given by
\small
\begin{align}
I_I&=G_NP_I\sum_{x\in\Phi_I}S_{xo}D_{xo}^{-\alpha}\nonumber\\
I_A&=\frac{G_NP_A}{N-1}\sum_{x\in\Phi_A}S_{xo}D_{xo}^{-\alpha}\label{eq:agg-int-inf-an-st}
\end{align}
\normalsize
where~$D_{xo}$ represents the distance from the transmitter at~$x$ to the typical receiver at the origin, and~$S_{xo}$ represents the corresponding channel gain. With independent Rayleigh fading, the channel gains~$S_{xo}$ are independent and exponentially distributed with unit mean, i.e.,~$S_{xo}\sim{\rm Exp}\left(1\right)$. The channel gain from the typical transmitter to the typical receiver is denoted by~$S_{o}$, with~$S_{o}\sim{\rm Exp}\left(1\right)$. The SIR at the typical receiver is given by
\begin{align}
{\rm SIR}_o=G_NP_IS_o{r^{-\alpha}}\left(I_I+I_A\right)^{-1}\label{eq:sir-pco-st}
\end{align}
and the connection outage probability is given by
\begin{align}
p_{{\rm co}}&=\Pr\left({\rm SIR}_{o}\leq\beta_{b}\right)\;.\label{eq:pco-as-gen}
\end{align}
This can be derived in closed-form and is presented in the following theorem:

\begin{PCO-ST}\label{thm:PCO-ST}
The connection outage probability of the sectoring scheme in~(\ref{eq:pco-as-gen}) is given by
\begin{align}
p_{{\rm co}}\!\!=\!1\!\!-\!\exp\!\left(\!{-\frac{\beta_{b}^{\frac{2}{\alpha}}\lambda_lC_{\alpha,2}r^{2}}{N}\!\!\left(\!1\!\!+\!\left(\!N\!-\!1\!\right)^{1\!-\!\frac{2}{\alpha}}\left(\!\phi^{-1}\!-\!1\!\right)^{\frac{2}{\alpha}}\!\right)}\!\!\right)\!\label{eq:pco-as}
\end{align}
where
\begin{align}
C_{\alpha,N}:=\pi\frac{\Gamma\left(N-1+\frac{2}{\alpha}\right)\Gamma\left(1-\frac{2}{\alpha}\right)}{\Gamma\left(N-1\right)}\label{eq:const-c-alpha-n}
\end{align}
and~$\Gamma\left(\cdot\right)$ is the gamma function.
\end{PCO-ST}
\begin{IEEEproof}
See Appendix~\ref{prf:PCO-ST}.
\end{IEEEproof}

From~(\ref{eq:pco-as}), we observe that as the distance between each transmitter-receiver pair~$r$ or the density of the transmitters~$\lambda_l$ increases, the connection outage probability increases towards one~(as expected), with the rate of increase being exponential in~$r^2$ or~$\lambda_l$. In the latter case, as~$\lambda_l$ increases, more transmitters ``on average'' will be closer to any given legitimate receiver, leading to an increased connection outage probability.

From~(\ref{eq:pco-as}), for a given power allocation ratio, it can be shown that with a relatively small path-loss exponent, i.e.,~$\alpha=2\sim4$, the connection outage probability decreases when extra directional antennas are added. The improvement comes from two aspects: 1) the intended sectors shrink and thus the legitimate receivers are interfered by less information signals; 2) the power allocated to the artificial noise is distributed in more sectors and thus the legitimate receivers are interfered by relatively less artificial noise.

\subsection{Secrecy Outage Probability}
Here, we characterize the secrecy outage probability~$p_{\rm so}$. A threshold SIR value for secrecy outage is defined as~$\beta_{e}$. As before, we focus on a typical transmitter-receiver pair but shift the coordinate system to put the transmitter at the origin. The message from the typical transmitter is not secure against the eavesdropper at~$z$ if~${\rm SIR}_z>\beta_e$, where~${\rm SIR}_{z}$ denotes the SIR received by the eavesdropper at~$z$. With the sectoring scheme, only the eavesdroppers inside the intended sector of the typical transmitter may cause secrecy outage. Those eavesdroppers form a fan-shaped PPP and by~\cite[Theorem~A.1]{Haenggi2009B}, they can be mapped as a homogeneous PPP on the whole plane, denoted by~$\Phi_{Z}$ with density~$\frac{1}{N}\lambda_{e}$.

As done in~\cite{Koyluoglu2010,Vasudevan2010,Zhou2012}, we design for a worst-case scenario by overestimating the eavesdroppers' multi-user decodability: The aggregate interference at the eavesdropper side consists of the artificial noise only. By~\cite{Haenggi2009B}, the transmitters which are radiating artificial noise towards the eavesdropper at~$z$ form a homogeneous PPP~$\Phi_A$ with density~$\frac{N-1}{N}\lambda_l$. Hence, the interference seen by the eavesdropper at~$z$ is
\begin{align}
I_A&=\frac{G_NP_A}{N-1}\sum_{x\in\Phi_A}S_{xz}D_{xz}^{-\alpha}\label{eq:agg-int-an-st}
\end{align}
where~$D_{xz}$ represents the distance from the transmitter at~$x$ to the eavesdropper at~$z$, whilst~$S_{xz}\sim{\rm Exp}\left(1\right)$ represents the corresponding channel gain.

The SIR received by the eavesdropper at~$z$ is given by
\begin{align}
{\rm SIR}_{z} & ={G_NP_IS_{oz}D_{oz}^{-\alpha}}{I_A^{-1}}\label{eq:sirz-st}
\end{align}
where~$D_{oz}$ represents the distance from the typical transmitter to the eavesdropper at~$z$, whilst~$S_{oz}\sim{\rm Exp}\left(1\right)$ represents the corresponding channel gain.

By taking the complement of the probability that the transmitted message is secure against all of the eavesdroppers, the secrecy outage probability can be expressed as
\begin{align}
p_{{\rm so}}&=1-\mathbb{E}_{\Phi_A}\left\{\mathbb{E}_{\Phi_{Z}}\left\{\prod_{z\in\Phi_{Z}}{\rm \Pr}\left({\rm SIR}_{z}<\beta_{e}|\Phi_A\right)\right\}\right\}\;.\label{eq:pso-as-ex}
\end{align}
The following theorem presents closed-form upper and lower bounds for this quantity:
\begin{PSO-ST}\label{thm:PSO-ST}
The secrecy outage probability of the sectoring scheme in~(\ref{eq:pso-as-ex}) satisfies
\begin{align}
p_{\rm so}^{\rm LB}\leq p_{\rm so}\leq p_{\rm so}^{\rm UB}
\end{align}
where
\begin{align}
p_{\rm so}^{\rm UB}
&=1-\exp\left(-\frac{\frac{\pi}{C_{\alpha,2}}\frac{\lambda_{e}}{\lambda_l}}{\beta_{e}^{\frac{2}{\alpha}}\left(N-1\right)^{1-\frac{2}{\alpha}}\left(\phi^{-1}-1\right)^{\frac{2}{\alpha}}}\right)\label{eq:pso-as-ub}
\end{align}
and
\begin{align}
p_{\rm so}^{\rm LB}=\frac{\pi\lambda_e}{\pi\lambda_e+{\lambda_lC_{\alpha,2}}\beta_e^{\frac{2}{\alpha}}\left(N-1\right)^{1-\frac{2}{\alpha}}\left(\phi^{-1}-1\right)^{\frac{2}{\alpha}}}\label{eq:pso-as-lb}
\end{align}
with~$C_{\alpha,2}$ defined in~(\ref{eq:const-c-alpha-n}).
\end{PSO-ST}
\begin{IEEEproof}
See Appendix~\ref{prf:PSO-ST}.
\end{IEEEproof}

\begin{figure}[t]
\centering
\includegraphics[width=0.9\linewidth]{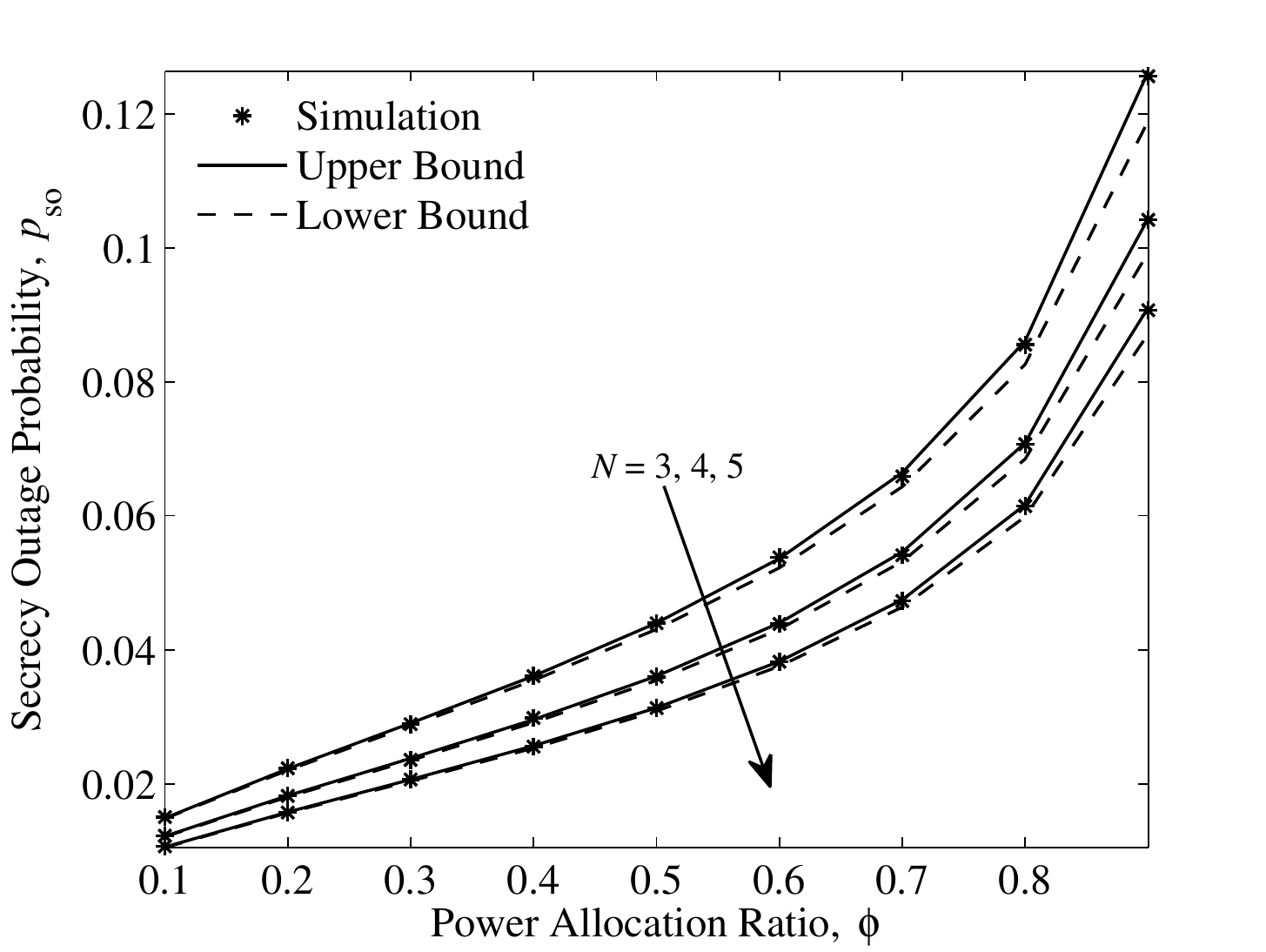}
\caption{The secrecy outage probability bounds of the sectoring scheme in~(\ref{eq:pso-as-ub}) and~(\ref{eq:pso-as-lb}) and the simulation results versus the power allocation ratio. Results are shown for the case where~$\alpha=4$,~$\lambda_l=0.01$,~$\lambda_e=0.001$ and~$\beta_e=1$.} \label{fig:pso_phi_n_ub_lb_st}
\end{figure}

In the low outage region~(i.e., as~$p_{\rm so}\to0$), the upper bound in~(\ref{eq:pso-as-ub}) and the lower bound in~(\ref{eq:pso-as-lb}) share the same leading order term:
\begin{align}
p_{\rm so}^{\rm UB}\approx\frac{\frac{\pi}{C_{\alpha,2}}\frac{\lambda_{e}}{\lambda_l}}{\beta_{e}^{\frac{2}{\alpha}}\left(N-1\right)^{1-\frac{2}{\alpha}}\left(\phi^{-1}-1\right)^{\frac{2}{\alpha}}}\approx p_{\rm so}^{\rm LB}
\end{align}
implying that they both approach the exact secrecy outage probability as it becomes
small. As shown in Fig.~\ref{fig:pso_phi_n_ub_lb_st}, the upper bound in~(\ref{eq:pso-as-ub}) gives a very accurate approximation for the entire range of the secrecy outage probabilities shown, while the lower bound in~(\ref{eq:pso-as-lb}) gets asymptotically accurate as the secrecy outage probability becomes small.

From~(\ref{eq:pso-as-ub}), we see that the secrecy outage probability increases with increasing the eavesdroppers' density~$\lambda_e$. Intuitively, as~$\lambda_e$ increases, more eavesdroppers ``on average'' will be closer to any given transmitter, which leads to an increased secrecy outage probability. More interestingly, the secrecy outage probability in~(\ref{eq:pso-as-ub}) depends on the densities of the transmitter-receiver pairs and the eavesdroppers solely through their ratio~$\frac{\lambda_e}{\lambda_l}$.

From~(\ref{eq:pso-as-ub}), we observe that for a given power allocation ratio, the secrecy outage probability can be reduced by adding extra directional transmit antennas. This result follows the intuition that by adding transmit antennas: 1) the intended sector shrinks and thus less eavesdroppers may cause secrecy outage; 2) the artificial noise from other transmitters covers a larger region and thus more eavesdroppers are degraded.

\section{Outage Performance of Beamforming Scheme}\label{sec:outage-beafroming}
In this section, we characterize the outage performance of the beamforming scheme. As before, we first derive the outage probability of the intended links. Then, we inspect the possibility that the transmitted message is not secure against malicious eavesdropping.

\subsection{Connection Outage Probability}
To facilitate our subsequent analysis, we first derive the distribution of the interference power resulting from the beamforming signal in~(\ref{eq:an-an}).

\begin{PCO-INT}\label{thm:PCO-INT}
For a given realization of the intended channel~${\bf h}$, if the transmitted signal~${\bf x}$ in~(\ref{eq:an-an}) is received by a non-intended receiver through an unknown channel~${\bf h}_z$, and if~$P_I\neq\frac{P_A}{N-1}$~(i.e.,~$\phi\neq\frac{1}{N}$), then the resulting interference power~$P_x$ is distributed according to the following probability density function~(p.d.f.):
\small
\begin{align}
f_{P_x}\left(z\right)&=\frac{1}{P_I}\left(1-\frac{P_A}{\left(N-1\right)P_I}\right)^{1-N}{\rm e}^{-\frac{z}{P_I}}\label{eq:pdf-pi-bf}\\
&\times\left(1-{\rm e}^{-\left(\frac{N-1}{P_A}-\frac{1}{P_I}\right)z}\sum_{k=0}^{N-2}\frac{\left(\frac{N-1}{P_A}-\frac{1}{P_I}\right)^kz^k}{k!}\right)\;,~z\!>\!0\;.\nonumber
\end{align}
\normalsize
Meanwhile, if~$\phi=\frac{1}{N}$,~$P_x$ is gamma distributed with shape parameter~$N$ and scale parameter~$P_I$, i.e.,~$P_x\sim{\rm Gamma}\left(N,P_I\right)$. Note that the distribution of~$P_x$ is independent of the intended channel~(i.e.,~${\bf h}$).
\end{PCO-INT}
\begin{IEEEproof}
See Appendix~\ref{prf:PCO-INT}.
\end{IEEEproof}

With Lemma~\ref{thm:PCO-INT}, we now derive the connection outage probability~$p_{\rm co}$. A threshold SIR value for connection outage is defined as~$\beta_{b}$. As before, we focus on a typical transmitter-receiver pair, and put the typical receiver at the origin to observe the network performance. We denote the interference power from the transmitter at~$x$ to the typical receiver by~$P_{xo}$, where the path loss is excluded. Note that~$P_{xo}$ admits the p.d.f. in~(\ref{eq:pdf-pi-bf}). The aggregate interference seen by the typical receiver is given by
\begin{align}
I_{IA}=\sum_{x\in\Phi_L}P_{xo}D_{xo}^{-\alpha}\label{eq:agg-int-an-all}
\end{align}
where~$D_{xo}$ denotes the distance from the transmitter at~$x$ to the typical receiver at the origin.

The channel vector from the typical transmitter to the typical receiver is denoted as~${\bf h}_o$. With Rayleigh fading, the elements of~${\bf h}_o$ are independent complex Gaussian distributed with zero mean and unit variance. With the beamforming strategy in~(\ref{eq:an-an}), the SIR at the typical receiver is given by
\begin{align}
{\rm SIR}_o={P_I\|{\bf h}_o\|^2r^{-\alpha}}{I_{IA}^{-1}}\label{eq:sir-pco-an}
\end{align}
and the connection outage probability is then given by
\begin{align}
p_{\rm co}=\Pr\left({\rm SIR}_o\leq\beta_b\right)\;.\label{eq:pco-ang-gen}
\end{align}
We have the following key theorem:
\begin{PCO-AN}\label{thm:PCO-AN}
The connection outage probability of the beamforming scheme in~(\ref{eq:pco-ang-gen}) admits
\small
\begin{align}
p_{\rm co}\!\!&=\!\!1\!\!-\!{\rm e}^{-{\beta_b^\frac{2}{\alpha}}\!\psi\left(\phi\right)}\!\!-\!{\rm e}^{-{\beta_b^\frac{2}{\alpha}}\!\psi\left(\phi\right)}\!\!\sum_{p=1}^{N-1}\!\frac{1}{p!}\!\sum_{k=1}^{p}\!\left(\!\frac{2}{\alpha}{\beta_b^\frac{2}{\alpha}}\!\psi\!\left(\phi\right)\!\right)^k\!\zeta\!\left(p,\!k\right)\label{eq:pco-an-ext-exp}
\end{align}
\normalsize
where~$\psi\left(\phi\right)$ is defined in~(\ref{eq:psi-phi-an}) and
\begin{figure*}[!t]
\begin{align}
\psi\left(\phi\right)&=\begin{cases}
\pi\lambda_lr^2\Gamma\left(1-\frac{2}{\alpha}\right)\left(\frac{N-\phi^{-1}}{N-1}\right)^{1-N}\left(\Gamma\left(1+\frac{2}{\alpha}\right)-\left(\frac{\phi^{-1}-1}{N-1}\right)^{1+\frac{2}{\alpha}}\sum_{k=0}^{N-2}\left(\frac{N-\phi^{-1}}{N-1}\right)^k\frac{\Gamma\left(k+1+\frac{2}{\alpha}\right)}{\Gamma\left(k+1\right)}\right)&{\rm if}~\phi\neq\frac{1}{N}\\
{\pi\lambda_lr^2}\frac{\Gamma\left(1-\frac{2}{\alpha}\right)\Gamma\left(N+\frac{2}{\alpha}\right)}{\Gamma\left(N\right)}&{\rm if}~\phi=\frac{1}{N}\\
\end{cases}\label{eq:psi-phi-an}
\end{align}
\normalsize \hrulefill
\end{figure*}
\begin{align}
\zeta\!\left(p,\!k\right)&\!=\!\sum_{\theta\in{\rm comb}\binom{p-1}{p-k}}\!\prod_{\substack{l_{i}\in\theta\\i=1,\ldots,p-k}}\!\left(\!l_{i}-\frac{2}{\alpha}\left(l_{i}-i+1\right)\!\right)\!\;.\label{eq:zeta-an}
\end{align}
Here~${\rm comb}\binom{p-1}{p-k}$ is the set of all distinct subsets of the natural
numbers~$\{1,2,\ldots,p-1\}$ with cardinality~$p-k$. For each subset, the elements are arranged in an increasing order and~$l_i\in\theta$ is the~$i$-th element of~$\theta$. For~$p\geq1$,~$\zeta\left(p,p\right)=1$.
\end{PCO-AN}
\begin{IEEEproof}
See Appendix~\ref{prf:PCO-AN}.
\end{IEEEproof}

Due to the complexity of the derived expression in~(\ref{eq:pco-an-ext-exp}), observing the effects of varying the key system parameters such as the power allocation ratio~$\phi$ and the number of transmit antennas~$N$ seems difficult. Nevertheless, with~(\ref{eq:pco-an-ext-exp}), we may readily evaluate the system performance and thereby avoid large-scale network simulations. In the low outage region, a much simpler approximation can be found as follows:
\begin{PCO-LOW-AN}\label{thm:PCO-LOW-AN}
In the low outage region~(i.e., as~$p_{\rm co}\to0$), the connection outage probability in~(\ref{eq:pco-an-ext-exp}) can be approximated by
\begin{align}
\tilde{p}_{\rm co}&=\beta_b^{\frac{2}{\alpha}}{\psi\left(\phi\right)}{K_{\alpha,N}}\label{eq:pco-an-app}
\end{align}
where~$\psi\left(\phi\right)$ is defined in~(\ref{eq:psi-phi-an}) and
\begin{align}
K_{\alpha,N}&=1-\frac{2}{\alpha}\sum_{p=1}^{N-1}\frac{1}{p!}\prod_{l=1}^{p-1}\left(l-\frac{2}{\alpha}\right)\;.\label{eq:kan-an}
\end{align}
\end{PCO-LOW-AN}
\begin{IEEEproof}
See Appendix~\ref{prf:PCO-LOW-AN}.
\end{IEEEproof}

\begin{figure}[t]
\centering
\includegraphics[width=0.9\linewidth]{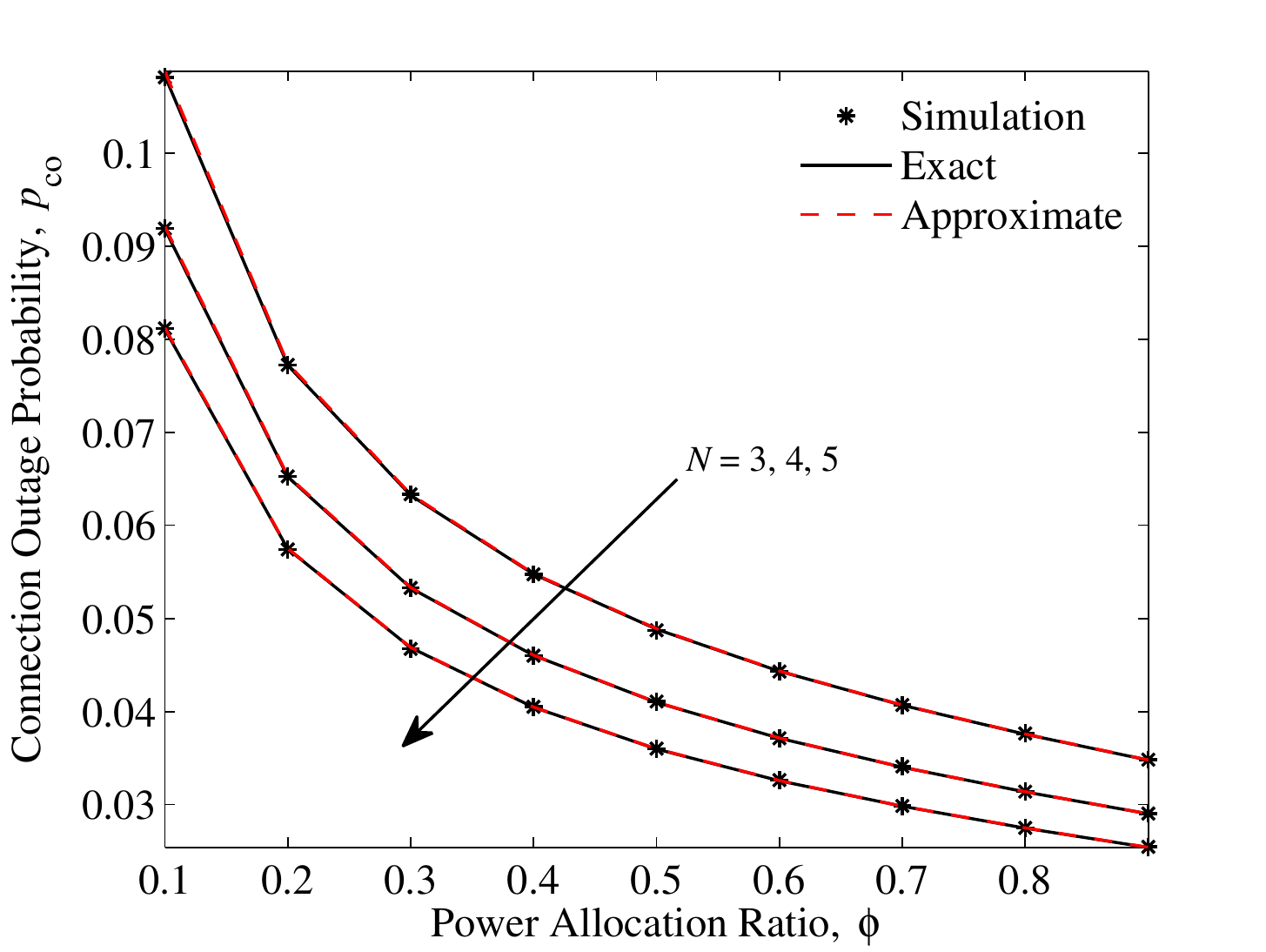}
\caption{The connection outage probability of the beamforming scheme in~(\ref{eq:pco-an-ext-exp}), the low outage approximation in~(\ref{eq:pco-an-app}) and the simulation results versus the power allocation ratio. Results are shown for the case where~$\lambda_l=0.01$, $r=1$,~$\alpha=4$ and~$\beta_b=3$.} \label{fig:pco_phi_n_ext_app_an}
\end{figure}

As demonstrated in Fig.~\ref{fig:pco_phi_n_ext_app_an}, for outage probabilities of practical interests, the difference between the exact value and the low outage approximation in~(\ref{eq:pco-an-app}) can hardly be seen.

\subsection{Secrecy Outage Probability}
Here we derive the secrecy outage probability~$p_{\rm so}$. A threshold SIR value for secrecy outage is defined as~$\beta_{e}$. As before, we focus on a typical transmitter-receiver pair but shift the coordinate system to put the transmitter at the origin.

As done in~\cite{Koyluoglu2010,Vasudevan2010,Zhou2012}, we design for a worst-case scenario by overestimating the eavesdroppers' multi-user decodability: The aggregate interference at the eavesdropper side consists of the artificial noise only. We denote the channel from the transmitter at~$x$ to the eavesdropper at~$z$ by~${\bf h}_{xz}$. For the eavesdropper at~$z$, by~(\ref{eq:an-an}), the artificial noise received from the transmitter at~$x$ is given by~${\bf h}_{xz}^H{\bf W}_x{\bf v}_x$, where~${\bf W}_x$ and~${\bf v}_x$ constitute the beamforming matrix and the artificial noise vector for the transmitter at~$x$. Define~${\bf g}_{xz}:={\bf h}_{xz}^H{\bf W}_x$ and note that~${\bf g}_{xz}$ is a~$N-1$ dimensional row vector. The corresponding interference power is given by~$\|{\bf g}_{xz}\|^{2}\sigma_v^2$, where~$\sigma_v^2=\frac{P_A}{N-1}$ is the variance of each element of~${\bf v}_x$. Then, for the eavesdropper at~$z$, the aggregate interference created by the artificial noise from the transmitters in~$\Phi_L$ is given by
\begin{align}
I_A=\frac{P_A}{N-1}\sum_{x\in\Phi_L}\|{\bf g}_{xz}\|^2D_{xz}^{-\alpha}\label{eq:agg-int-an}
\end{align}
where~$D_{xz}$ denotes the distance from the transmitter at~$x$ to the eavesdropper at~$z$. Since the columns of~${\bf W}_x$ are unit-norm and mutually orthogonal, the elements of~${\bf g}_{xz}$ are independent complex Gaussian distributed with zero mean and unit variance; thus, we know that~$\|{\bf g}_{xz}\|^2\sim{\rm Gamma}\left(N-1,1\right)$. Note that the eavesdroppers will also be jammed by the artificial noise from the typical transmitter. Similar to the discussions above, for the eavesdropper at~$z$, the interference power created by the typical transmitter is given by~$\frac{P_A}{N-1}\|{\bf g}_{oz}\|^2D_{oz}^{-\alpha}$, where~$\|{\bf g}_{oz}\|^2\sim{\rm Gamma}\left(N-1,1\right)$, and~$D_{oz}$ is the distance from the typical transmitter to the eavesdropper at~$z$.

Denote~${\bf h}_o$ as the channel from the typical transmitter to the typical receiver, and~${\bf h}_{oz}$ as the channel from the typical transmitter to the eavesdropper at~$z$. Then, denote the channel gain resulting from Rayleigh fading from the typical transmitter to the eavesdropper at~$z$ by~$S_{oz}$. Due to the fact that the typical transmitter is beamforming towards the typical receiver, the channel gain~$S_{oz}$ is given by
\begin{align}
S_{oz}=\left|{\bf h}_{oz}^H\frac{{\bf h}_o}{\|{\bf h}_o\|}\right|^2\sim{\rm Exp}\left(1\right)\;.
\end{align}
The SIR for the eavesdropper at~$z$ is then given by
\begin{align}
{\rm SIR}_z=\frac{P_IS_{oz}D_{oz}^{-\alpha}}{\frac{P_A}{N-1}\|{\bf g}_{oz}\|^2D_{oz}^{-\alpha}+I_A}\;.\label{eq:sirz-an}
\end{align}

By taking the complement of the probability that the transmitted message is secure against all of the eavesdroppers, the secrecy outage probability can be expressed as
\begin{align}
p_{\rm so}\!=\!1\!-\!\mathbb{E}_{\Phi_L}\!\left\{\mathbb{E}_{\Phi_E}\!\left\{\prod_{z\in\Phi_E}\Pr\left({\rm SIR}_z\!<\!\beta_e|\Phi_L\right)\right\}\!\right\}\!\;.\label{eq:pso-gen-1}
\end{align}
We can find closed-form upper and lower bounds as follows:

\begin{PSO-AN}\label{thm:PSO-AN}
The secrecy outage probability of the beamforming scheme in~(\ref{eq:pso-gen-1}) satisfies
\begin{align}
p_{\rm so}^{\rm LB}\leq p_{\rm so}\leq p_{\rm so}^{\rm UB}
\end{align}
where
\begin{align}
p_{\rm so}^{\rm UB}
=1-\exp\left[-\frac{\lambda_e}{\lambda_l}\frac{\pi\left(\beta_e\frac{\phi^{-1}-1}{N-1}+1\right)^{1-N}}{C_{\alpha,N}\left(\beta_e\frac{\phi^{-1}-1}{N-1}\right)^{\frac{2}{\alpha}}}\right]\label{eq:pso-an-ub}
\end{align}
and
\begin{align}
p_{\rm so}^{\rm LB}\!=\!\left(\beta_e\frac{\phi^{-1}\!-\!1}{N\!-\!1}\!+\!1\right)^{1\!-\!N}\!\frac{\pi\lambda_e}{\pi\lambda_e\!+\!\lambda_lC_{\alpha,N}\!\left(\!\beta_e\frac{\phi^{-1}-1}{N-1}\!\right)^{\frac{2}{\alpha}}}\label{eq:pso-an-lb}
\end{align}
with~$C_{\alpha,N}$ defined in~(\ref{eq:const-c-alpha-n}).
\end{PSO-AN}
\begin{IEEEproof}
See Appendix~\ref{prf:PSO-AN}.
\end{IEEEproof}

\begin{figure}[t]
\centering
\includegraphics[width=0.9\linewidth]{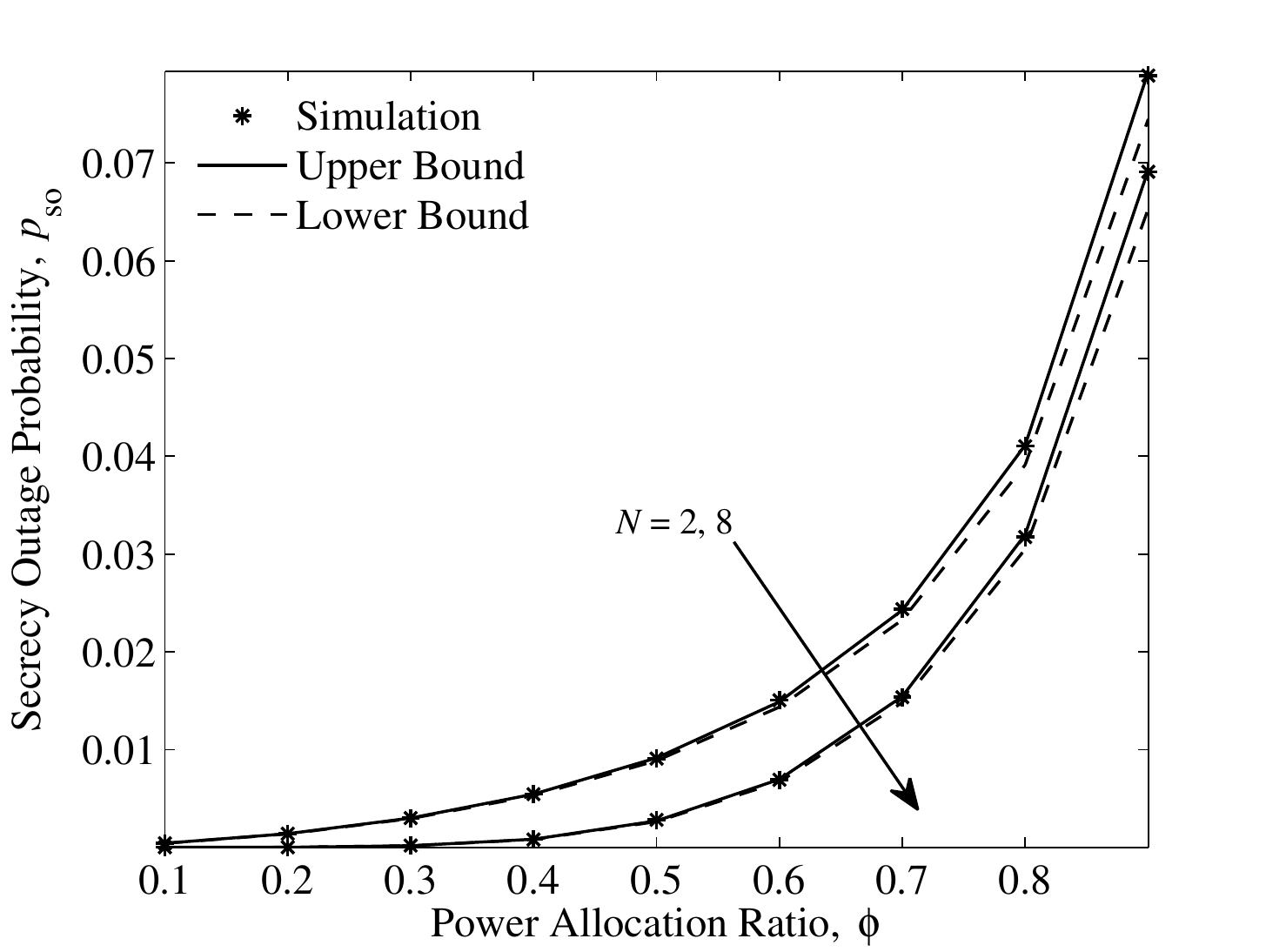}
\caption{The secrecy outage probability bounds of the beamforming scheme in~(\ref{eq:pso-an-ub}) and~(\ref{eq:pso-an-lb}) and the simulation results versus the power allocation ratio. Results are shown for the case where~$\alpha=4$,~$\lambda_l=0.01$,~$\lambda_e=0.001$ and~$\beta_e=3$.} \label{fig:pso_phi_n_ub_lb_an}
\end{figure}

In the low outage region~(i.e., as~$p_{\rm so}\to0$), the upper bound in~(\ref{eq:pso-an-ub}) and the lower bound in~(\ref{eq:pso-an-lb}) share the same leading order term:
\begin{align}
p_{\rm so}^{\rm UB}\approx p_{\rm so}^{\rm Lead}:=\frac{\lambda_e}{\lambda_l}\frac{\pi\left(\beta_e\frac{\phi^{-1}-1}{N-1}+1\right)^{1-N}}{C_{\alpha,N}\left(\beta_e\frac{\phi^{-1}-1}{N-1}\right)^{\frac{2}{\alpha}}}\approx p_{\rm so}^{\rm LB}\label{eq:leading-pso-an}
\end{align}
implying that these two bounds both capture the leading order behavior of the secrecy outage probability as it becomes small. The leading order term~$p_{\rm so}^{\rm Lead}$ can serve as an accurate approximation for the secrecy outage probability in the low outage region. As shown in Fig.~\ref{fig:pso_phi_n_ub_lb_an}, the upper bound in~(\ref{eq:pso-an-ub}) gives a very accurate approximation for the entire range of the secrecy outage probabilities shown, while the lower bound in~(\ref{eq:pso-an-lb}) gets asymptotically accurate as the secrecy outage probability becomes small.

We point out that with the p.d.f. provided in Lemma~\ref{thm:PCO-INT}, following a similar procedure as that used in deriving Theorem~\ref{thm:PSO-AN}, it is not difficult to study the case where the eavesdroppers do not have multi-user decodability and simply treat interference as noise. This argument also applies to the sectoring scheme. However, as we mentioned in the introduction, in order to achieve the maximum level of secrecy, assuming eavesdroppers with multi-user decodability is a more robust approach. Therefore, in this paper, we retain our focus on this case.

\begin{figure}[t]
\centering
\includegraphics[width=0.9\linewidth]{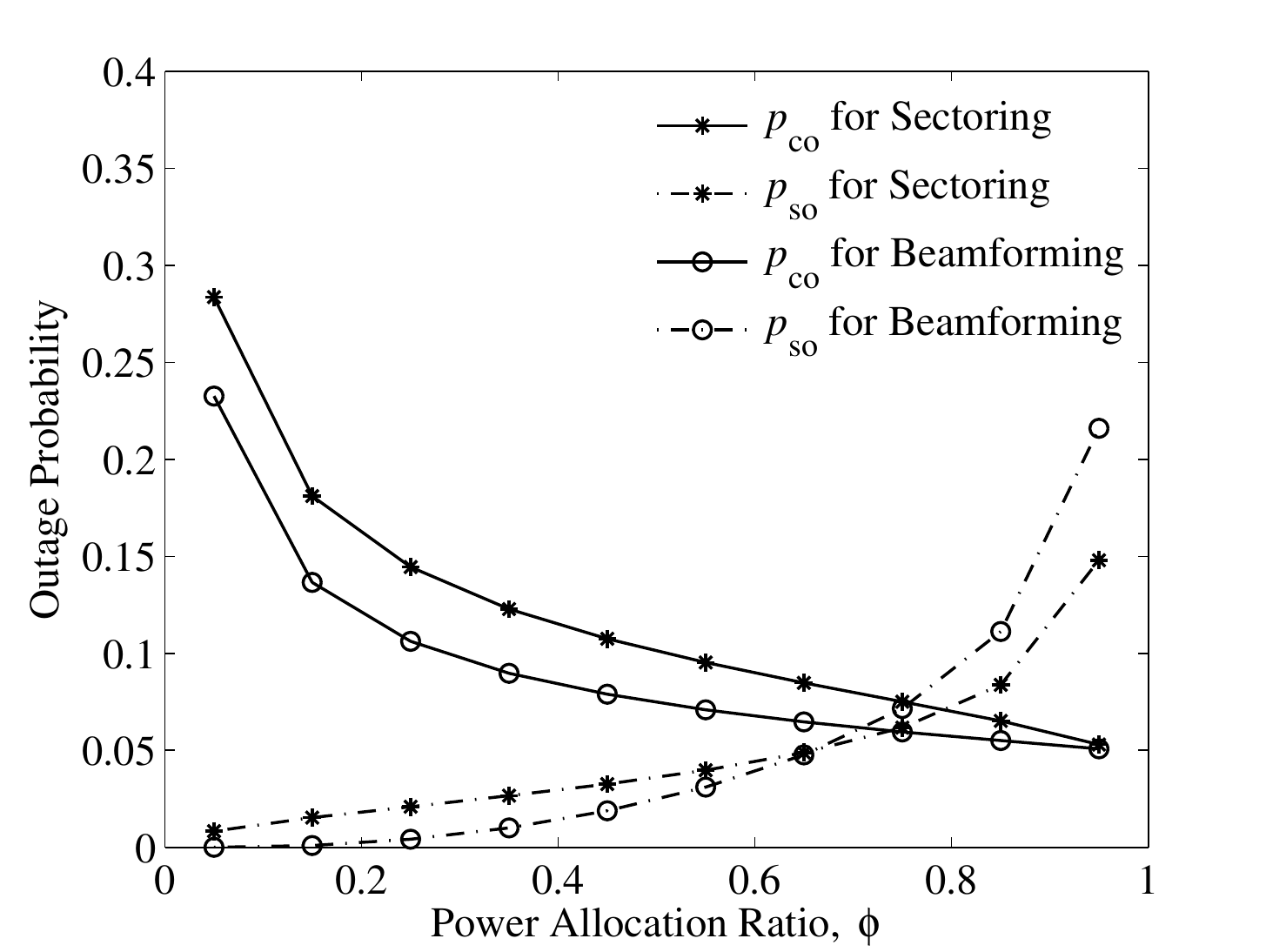}
\caption{The outage probabilities of the sectoring and beamforming schemes in~(\ref{eq:pco-as}),~(\ref{eq:pso-as-ub}),~(\ref{eq:pco-an-ext-exp}) and~(\ref{eq:pso-an-ub}) versus the power allocation ratio. Results are shown for the case where~$\lambda_l=0.01$,~$\lambda_e=0.001$,~$r=1$,~$\alpha=4$,~$N=4$,~$\beta_b=10$ and~$\beta_e=1$.} \label{fig:outage_prob_st_an}
\end{figure}

\underline{{\bf Connection and Secrecy Outage Tradeoff}}: As shown in Fig.~\ref{fig:outage_prob_st_an}, for both the sectoring and beamforming schemes, when giving more power to the information signal~(i.e., increasing the power allocation ratio~$\phi$), the connection outage probability~$p_{\rm co}$ decreases whilst the secrecy outage probability~$p_{\rm so}$ increases. Hence, there is a tradeoff between the connection and secrecy outage performance w.r.t. the transmit power allocation.

\section{Secrecy Throughput Performance}\label{sec:throughput}
In this section, based on the outage probability expressions obtained in the last two sections, we investigate the networkwide secrecy throughput performance for the sectoring and beamforming schemes, in terms of the secrecy transmission capacity~\cite{Zhou2011B}.

The secrecy transmission capacity is defined as the achievable rate of successful transmission of
confidential messages per unit area with constraints on both the connection and secrecy outage probabilities. With outage constraints~$p_{\rm co}=\sigma$ and~$p_{\rm so}=\epsilon$, the corresponding secrecy transmission capacity is given by
\begin{align}
C&=\left(1-\sigma\right)\lambda_lR_s\nonumber\\
&=\left(1-\sigma\right)\lambda_l\left[R_b-R_e\right]^+\label{eq:stc-def-as}
\end{align}
where~$\left[x\right]^+=\max\left\{0,x\right\}$. Note that the message rate~$R_s$ is a function of~$\sigma$ and~$\epsilon$, since~$R_b=\log_2\left(1+\beta_b\right)$ is determined by the connection outage constraint~$\sigma$, while~$R_e=\log_2\left(1+\beta_e\right)$ is determined by the secrecy outage constraint~$\epsilon$. Whenever~$R_b-R_e$ is negative, the required secrecy and connection outage performance cannot be guaranteed simultaneously, and the message transmission should be suspended. More discussions on the relationship between the transmission rates and the outage events can be found in Section~\ref{sec:code-outage}.

\subsection{Sectoring Scheme}
Here we characterize the secrecy transmission capacity of the sectoring scheme in the following proposition:
\begin{STC-ST}\label{thm:STC-ST}
For the sectoring scheme, a tight lower bound to the secrecy transmission capacity in~(\ref{eq:stc-def-as}) can be given by
\begin{align}
C_{\rm Sector}^{\rm LB}\!&=\!\left(1\!-\!\sigma\right)\!\lambda_l\label{eq:stc-lb-st}\\
&\times\left[\!\log_{2}\!\left(\!\frac{1+\left(\frac{\frac{N}{\lambda_lC_{\alpha,2}\,r^{2}}\ln\left(\frac{1}{1-\sigma}\right)}{1+\left(N-1\right)^{1-\frac{2}{\alpha}}\left(\phi^{-1}-1\right)^{\frac{2}{\alpha}}}\right)^{\frac{\alpha}{2}}}{1+\left(\frac{\frac{\pi}{C_{\alpha,2}}\frac{\lambda_{e}}{\lambda_l}}{\ln\left(\frac{1}{1-\epsilon}\right)\left(N-1\right)^{1-\frac{2}{\alpha}}\left(\phi^{-1}-1\right)^{\frac{2}{\alpha}}}\right)^{\frac{\alpha}{2}}}\!\right)\!\right]^{+}\nonumber
\end{align}
with~$C_{\alpha,2}$ defined in~(\ref{eq:const-c-alpha-n}).
\end{STC-ST}
\begin{IEEEproof}
From~(\ref{eq:pco-as}) and~(\ref{eq:pso-as-ub}), we solve~$p_{\rm co}=\sigma$ and~$p_{\rm so}^{\rm UB}=\epsilon$ w.r.t.~$R_b=\log_2\left(1+\beta_b\right)$ and~$R_e^{\rm UB}=\log_2\left(1+\beta_e\right)$, respectively. Then, plugging the obtained $R_b$ and $R_e^{\rm UB}$ into~(\ref{eq:stc-def-as}), we have the lower bound in~(\ref{eq:stc-lb-st}).
\end{IEEEproof}

Note that~(\ref{eq:stc-lb-st}) is derived based on the secrecy outage probability upper bound in~(\ref{eq:pso-as-ub}). Since~(\ref{eq:pso-as-ub}) tracks the exact secrecy outage probability very closely, the lower bound in~(\ref{eq:stc-lb-st}) provides a tight approximation to the actual secrecy transmission capacity. From~(\ref{eq:stc-lb-st}), we observe that the secrecy transmission capacity increases logarithmically as the number of transmit antennas grows large. The underlying reason is that under the outage constraints, as the number of transmit antennas grows large, the supported codeword rate~$R_b$ increases logarithmically, while the required rate redundancy~$R_e$ diminishes.

By properly adjusting the power allocation ratio~$\phi$, we can maximize the secrecy transmission capacity. Though a general expression for the optimal~$\phi$ seems not available, we still have the following corollary:
\begin{PHI-OPT-ST}\label{thm:PHI-OPT-ST}
The optimal power allocation ratio~$\phi^{*}$ of the sectoring scheme, which maximizes the tight secrecy transmission capacity lower bound~$C_{\rm Sector}^{\rm LB}$ in~(\ref{eq:stc-lb-st}), is unique and can be found by numerically solving for the root of the first-order derivative of~$C_{\rm Sector}^{\rm LB}$ w.r.t.~$\phi$. This is true even if the objective function is generally non-concave.
\end{PHI-OPT-ST}
\begin{IEEEproof}
See Appendix~\ref{prf:PHI-OPT-ST}.
\end{IEEEproof}

For the case where~$\alpha=4$, setting the derivative of~$C_{\rm Sector}^{\rm LB}$ in~(\ref{eq:stc-lb-st}) w.r.t.~$\phi$ to zero gives a cubic equation and solving it provides a closed-form expression for~$\phi^*$. Define:
\begin{align}
\varrho&\!=\!\frac{N}{\lambda_lC_{\alpha,2}\,r^{2}}\!\ln\!\left(\frac{1}{1-\sigma}\right)\;,~\varsigma\!=\!\frac{1}{\ln\!\left(\frac{1}{1-\epsilon}\right)}\frac{\pi}{C_{\alpha,2}}\frac{\lambda_{e}}{\lambda_l}\label{eq:phi-opt-a4-st-notations}\\
\kappa&\!=\!{\varrho}^{2}\!\!+\!{\varsigma}^{2}\!\!+\!\left(\!{\varrho}^{2}\!-\!{\varsigma}^{2}\!+\!\sqrt{\!\left(\!\left(\!\varrho\!-\!\varsigma\!\right)^{2}\!+\!1\!\right)\!\left(\!\left(\!\varrho\!+\!\varsigma\!\right)^{2}\!+\!1\!\right)\!}\!\right)\!\left(\!{\varrho}^{2}\!\!-\!{\varsigma}^{2}\!\right)\!\;.\nonumber
\end{align}
Then, the optimal power allocation ratio for~$\alpha=4$ is
\begin{align}
{\phi}^{\alpha=4}_{\rm Sector}\!\!=\!\!\left(\!\!1\!+\!\frac{\varsigma^{\frac{2}{3}}\!\!\left(\!{2}^{\frac{2}{3}}\!{\varrho}^{\frac{4}{3}}\!\varsigma^{\frac{1}{3}}\!{\kappa}^{\frac{2}{3}}\!+\!2^{\frac{4}{3}}\!{\varrho}^{2}\!\varsigma\!+\!2{\varrho}^{\frac{2}{3}}\!{\varsigma}^{\frac{5}{3}}\!{\kappa^{\frac{1}{3}}}\!\right)^2}{4\!\left(\!N\!-\!1\!\right)\!{\varrho}^{\frac{4}{3}}{\kappa^{\frac{2}{3}}}\!\left(\!{\varrho}^{2}\!-\!{\varsigma}^{2}\!\right)^2}\!\right)^{-1}\;.\label{eq:phi-opt-a4-st}
\end{align}

\begin{figure}[t]
\centering
\includegraphics[width=0.9\linewidth]{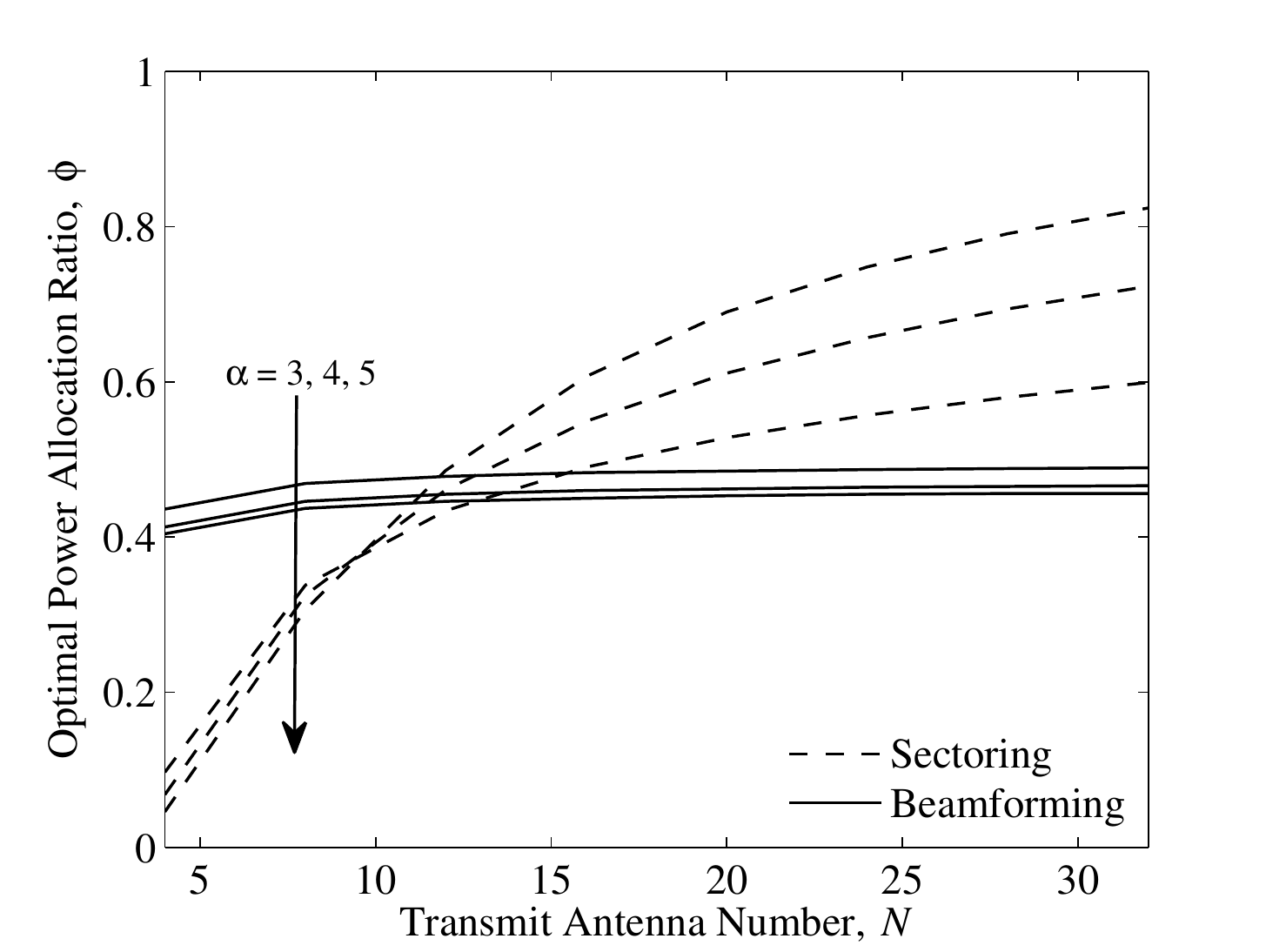}
\caption{The optimal power allocation ratio of the sectoring and beamforming schemes versus the number of transmit antennas. Results are shown for the case where~$\lambda_l=0.01$,~$\lambda_e=0.001$,~$r=1$,~$\sigma=0.1$ and~$\epsilon=0.01$.} \label{fig:phi_opt_n_alpha_st_an}
\end{figure}

We observe that as~$N\to\infty$,~${\phi}^{\alpha=4}_{\rm Sector}={1-\mathcal{O}\left(N^{-1}\right)}$. In other words, the optimal power allocation ratio increases towards one as the number of transmit antennas grows large. This observation can be explained by noting that adding extra transmit antennas allows the transmitter to concentrate more on the transmission towards the intended receiver, while less and less eavesdroppers may cause secrecy outage. Hence, more transmit power can be given to the information signal to achieve a better throughput performance, while still satisfying the connection and secrecy outage constraints. As can be seen from Fig.~\ref{fig:phi_opt_n_alpha_st_an}, the observations made from the case where~$\alpha=4$ holds more generally for~$3\leq\alpha\leq5$. The numerical results in Fig.~\ref{fig:stc_opt_n_alpha_st_an} indicate that with optimized power allocation, the secrecy transmission capacity increases logarithmically as the number of transmit antennas grows large, which agrees with our earlier observation made from~(\ref{eq:stc-lb-st}).

\subsection{Beamforming Scheme}
Now we study the secrecy transmission capacity of the beamforming scheme. Generally speaking, the secrecy transmission capacity can be computed by solving~$p_{\rm co}=\sigma$ w.r.t.~$R_b=\log_2\left(1+\beta_b\right)$ from~(\ref{eq:pco-an-ext-exp}) and solving~$p_{\rm so}=\epsilon$ w.r.t.~$R_e=\log_2\left(1+\beta_e\right)$ from~(\ref{eq:pso-gen-1}), and then plugging the obtained results into~(\ref{eq:stc-def-as}). However, the equation~$p_{\rm co}=\sigma$ seems not analytically solvable due to the existence of multiple summations. Note that the typical value of the connection outage constraint~$\sigma$ is expected to be small~(e.g., below~$0.1$). This allows us to use the low outage approximation provided in~(\ref{eq:pco-an-app}). Letting~$\tilde{p}_{\rm co}=\sigma$, recalling that~$R_b=\log_2\left(1+\beta_b\right)$, the supported codeword rate~$R_b$ can be approximated by
\begin{align}
\tilde{R}_b&=\log_2\left[1+\left(\frac{\sigma}{\psi\left(\phi\right)K_{\alpha,N}}\right)^{\frac{\alpha}{2}}\right]\label{eq:rb-app-an}
\end{align}
where~$\psi\left(\phi\right)$ and~$K_{\alpha,N}$ are defined in~(\ref{eq:psi-phi-an}) and~(\ref{eq:kan-an}), respectively.

Then, we need to solve~$p_{\rm so}=\epsilon$ w.r.t.~$R_e=\log_2\left(1+\beta_e\right)$. Note that the secrecy outage constraint~$\epsilon$ is also expected to be small~(e.g., below~$0.1$), this allows us to use the leading-order low outage approximation of the secrecy outage probability in~(\ref{eq:leading-pso-an}). In the special case of~$\alpha=4$ and~$N=2$, by solving~$p_{\rm so}^{\rm Lead}=\epsilon$, which turns to be a cubic equation, an approximation to the required rate redundancy~$R_e$ can be obtained as follows:
\small
\begin{align}
\tilde{R}_e\!\!=\!\log_2\!\!\left[\!1\!\!+\!\!\frac{\phi}{1\!-\!\phi}\!\left(\!\!\frac{\!\left(\!\frac{108\pi\lambda_e}{\epsilon\lambda_lC_{\alpha,2}}\!+\!12\sqrt{\!\left(\!\frac{9\pi\lambda_e}{\epsilon\lambda_lC_{\alpha,2}}\!\right)^{2}\!\!\!+\!12}\!\right)^{\frac{2}{3}}\!\!\!\!\!-\!12}{6\sqrt[3]{\frac{108\pi\lambda_e}{\epsilon\lambda_lC_{\alpha,2}}\!+\!\!12\sqrt{\!\left(\!\frac{9\pi\lambda_e}{\epsilon\lambda_lC_{\alpha,2}}\!\right)^{2}\!\!\!+\!\!12}}}\!\!\right)^2\right]\!\label{eq:a4-n2-re-an}
\end{align}
\normalsize
where~$C_{\alpha,2}$ is defined in~(\ref{eq:const-c-alpha-n}). Plugging~(\ref{eq:rb-app-an}) and~(\ref{eq:a4-n2-re-an}) into~(\ref{eq:stc-def-as}), we have a closed-form approximation for the secrecy transmission capacity of the beamforming scheme when~$\alpha=4$ and~$N=2$:
\begin{align}
C_{\rm Beam}^{\rm Approx}&=\left(1-\sigma\right)\lambda_l\left[\tilde{R}_{b}-\tilde{R}_{e}\right]^+\;.\label{eq:stc-asym-an}
\end{align}

Unfortunately, it seems that we cannot extend the analytical results to the case where~$N\geq3$, due to the fact that there is not a general algebraic solution for the quintic~(i.e., fifth order) equation and above~(see Abel's impossibility theorem in~\cite{Abel1826}, an alternative proof to it written in English can be found in~\cite{Zoladek2000}). Hence, for these cases, we study the secrecy transmission capacity numerically. From~(\ref{eq:leading-pso-an}), by solving~$p_{\rm so}^{\rm Lead}=\epsilon$ w.r.t.~$R_e=\log_2\left(1+\beta_e\right)$, an approximation to the required rate redundancy can be obtained and we denote it as~$\tilde{R}_e$. Then, plugging~$\tilde{R}_b$ in~(\ref{eq:rb-app-an}) and~$\tilde{R}_e$ into~(\ref{eq:stc-def-as}), we have an approximation for the secrecy transmission capacity of the beamforming scheme, which can be written as~(\ref{eq:stc-asym-an}). Note that this approximation is very accurate for small outage constraints~(i.e.,~$\sigma$ and~$\epsilon$), since the outage probability approximations in~(\ref{eq:pco-an-app}) and~(\ref{eq:leading-pso-an}) are both very accurate in the low outage region. We then numerically optimize the power allocation to maximize the secrecy transmission capacity.

\begin{figure}[t]
\centering
\includegraphics[width=0.9\linewidth]{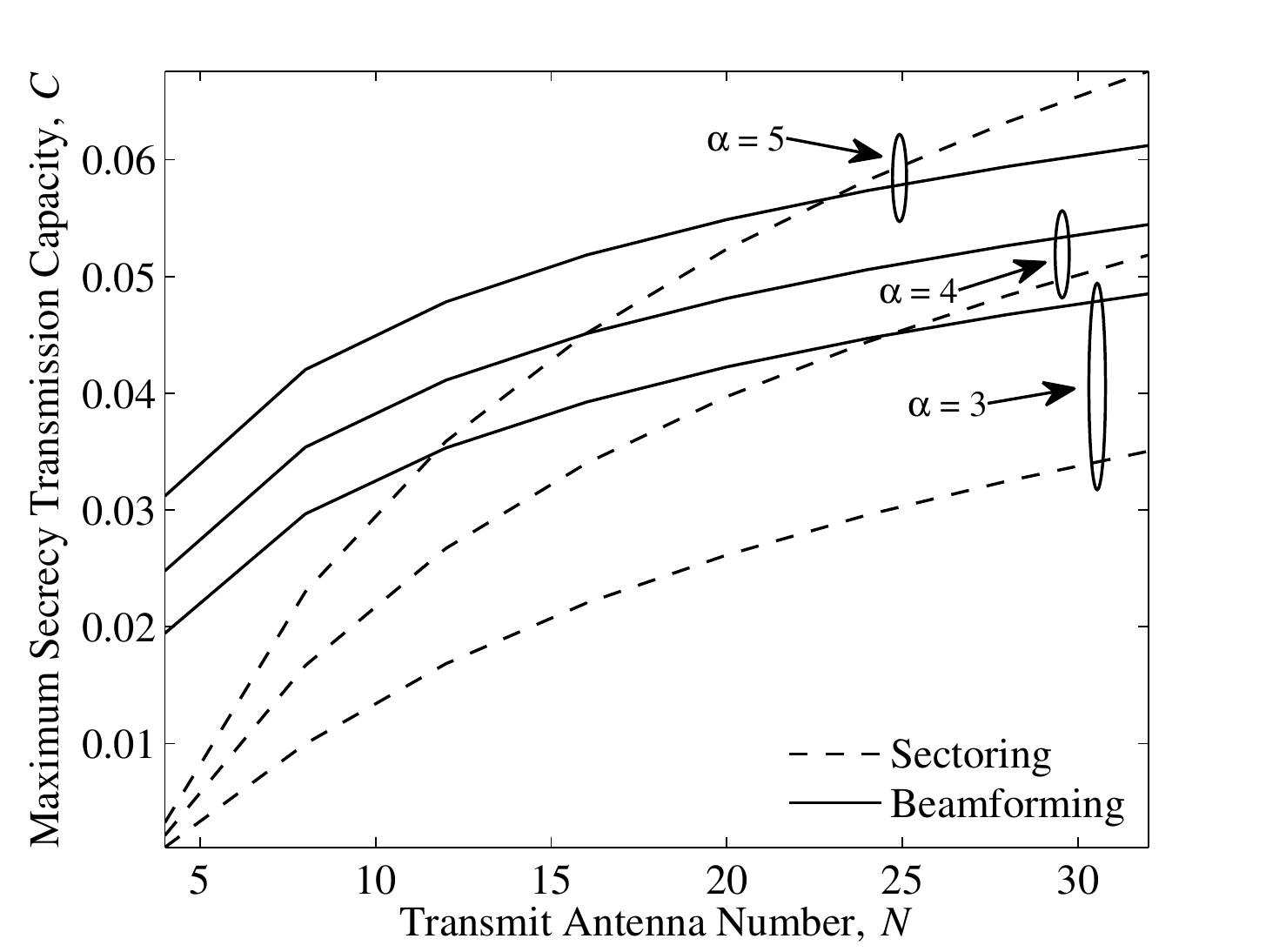}
\caption{The maximum secrecy transmission capacity of the sectoring and beamforming schemes versus the number of transmit antennas. Results are shown for the case where~$\lambda_l=0.01$,~$\lambda_e=0.001$,~$r=1$,~$\sigma=0.1$ and $\epsilon=0.01$.} \label{fig:stc_opt_n_alpha_st_an}
\end{figure}

As illustrated in Fig.~\ref{fig:phi_opt_n_alpha_st_an}, as the number of transmit antennas~$N$ grows large, the optimal power allocation ratio that maximizes the secrecy transmission capacity of the beamforming scheme converges to a certain value, which is strictly less than one. Here we explain why the optimal power allocation ratio does not increase to one.
Note that the density of the eavesdroppers which may cause secrecy outage does not change with increasing~$N$. More importantly, as~$N$ increases, the power of the emitted artificial noise decreases, and the increments of the artificial interference brought by adding extra transmit antennas will be neutralized. In such cases, giving too much power to the information signal will break the secrecy outage constraint and thus the optimal power allocation ratio will not increase to one. As demonstrated in Fig.~\ref{fig:stc_opt_n_alpha_st_an}, with optimized transmit power allocation, as~$N$ grows large, the maximum secrecy transmission capacity of the beamforming scheme increases logarithmically, which is similar to the sectoring scheme.

\subsection{Sectoring versus Beamforming}
Here we compare the sectoring and beamforming schemes, in terms of the optimal power allocation ratio that maximizes the secrecy transmission capacity and the corresponding maximum secrecy transmission capacity.

Regarding the optimal power allocation ratio:
\begin{itemize}
  \item As shown in Fig.~\ref{fig:phi_opt_n_alpha_st_an}, as the number of transmit antennas~$N$ increases, the optimal power allocation ratio for the beamforming scheme converges to a certain value which is less than one, while that for the sectoring scheme keeps increasing towards one. This difference can be explained by noting that for the beamforming scheme, with increasing~$N$, the equivalent density of eavesdroppers which may cause secrecy outage remains the same, while that for the sectoring scheme decreases inversely linearly. Hence, the reduced secrecy outage probability of the sectoring scheme allows the transmitters to give more transmit power to the information signal, while still satisfying the secrecy outage constraint, which is not true for the beamforming scheme.
  \item As shown in Fig.~\ref{fig:phi_opt_n_alpha_st_an}, when the path-loss becomes more severe~(i.e., as~$\alpha$ increases), the optimal power allocation ratio for the beamforming scheme decreases. More interestingly, with increasing~$\alpha$ and~$N$, there is twist in the optimal power allocation ratio of the sectoring scheme. The underlying reason is that with increasing~$N$, the optimal power allocation ratio of the sectoring scheme increases faster in a high path-loss environment.
\end{itemize}

Regarding the maximum secrecy transmission capacity:
\begin{itemize}
  \item As shown in Fig.~\ref{fig:stc_opt_n_alpha_st_an}, the beamforming scheme outperforms the sectoring scheme for a wide range of antenna numbers, in terms of achieving a larger secrecy transmission capacity. This makes sense because that having more channel knowledge at the transmitters, such that beamforming can be performed, is generally better than having directional transmit antennas, which are needed for the sectoring scheme.
  \item However, when~$N$ becomes sufficiently large, the sectoring scheme can achieve a larger secrecy transmission capacity, compared with the beamforming scheme. The reason behind this is that with increasing~$N$, the transmitter of the sectoring scheme becomes quite capable of concentrating the transmission towards the intended receiver while less and less eavesdroppers may cause secrecy outage. As can be seen from Fig.~\ref{fig:phi_opt_n_alpha_st_an}, the corresponding optimal power allocation ratio also increases with increasing~$N$, giving more transmit power to the information signals to achieve a better throughput performance. In addition, the performance crossover between the sectoring and beamforming schemes happens at a larger~$N$ if the secrecy outage constraint becomes more stringent~(i.e., reducing~$\epsilon$).
\end{itemize}

\section{Conclusion}\label{sec:conclusion}
In this paper, we studied physical-layer security in wireless ad hoc networks and investigated two types of multi-antenna transmission schemes. In particular, in order to jam the eavesdroppers over the entire network, we combined artificial noise with either sectoring or beamforming. For these two schemes, we provided closed-form expressions for the connection and secrecy outage probabilities and showed the tradeoff between them. We then quantified the secrecy throughput performance for both schemes in terms of the secrecy transmission capacity. Our results indicated that the proposed transmission schemes can provide significant secrecy enhancements over single-antenna methods. Our analysis also shed light on the behavior of the optimal power allocation between the information signal and the artificial noise for achieving the maximum secrecy throughput.

\appendix
\subsection{Proof of Theorem~\ref{thm:PCO-ST}}\label{prf:PCO-ST}
Here we derive the connection outage probability in~(\ref{eq:pco-as}). By~(\ref{eq:agg-int-inf-an-st}), the total interference seen by the typical receiver at the origin is~$I_I+I_A$. Since~$I_I$ and~$I_A$ are two independent shot noise processes, by~\cite[eq.~(8)]{Haenggi2009A}, the Laplace transform of the p.d.f. of~$I_I+I_A$ is given by
\begin{align}
\mathcal{L}_{I_I\!+\!I_A}\!\left(s\right)\!=\!\exp\!\left(\!\!-\frac{\lambda_lC_{\alpha,2}G_N^\frac{2}{\alpha}}{N}\!\!\left(\!P_I^{\frac{2}{\alpha}}\!\!+\!\left(\!N\!-\!1\!\right)^{1\!-\!\frac{2}{\alpha}}\!\!P_A^{\frac{2}{\alpha}}\!\right)\!s^{\frac{2}{\alpha}}\!\!\right)\!\label{eq:lap-agg-int-inf-an-st}
\end{align}
where~$C_{\alpha,2}$ is defined in~(\ref{eq:const-c-alpha-n}).

By~(\ref{eq:sir-pco-st}) and~(\ref{eq:pco-as-gen}), we have
\begin{align}
p_{\rm co}&=1-\Pr\left(S_{o}\geq\frac{\beta_{b}r^{\alpha}}{G_NP_I}\left(I_I+I_A\right)\right)\nonumber\\
&=1-\int_{0}^{\infty}\exp\left(-\frac{\beta_{b}r^{\alpha}}{G_NP_I}z\right)f_{I_I+I_A}\left(z\right){\rm d}z\nonumber\\
&=1-\mathcal{L}_{I_I+I_A}\left(\frac{\beta_{b}r^{\alpha}}{G_NP_I}\right)
\end{align}
where~$f_{I_I+I_A}\left(\cdot\right)$ is the p.d.f. of~$I_I+I_A$ and the last line comes from the definition of the Laplace transform. Then, plugging in the Laplace transform in~(\ref{eq:lap-agg-int-inf-an-st}) leads to the results in~(\ref{eq:pco-as}).

\subsection{Proof of Theorem~\ref{thm:PSO-ST}}\label{prf:PSO-ST}
We first derive the secrecy outage probability upper bound in~(\ref{eq:pso-as-ub}). By~\cite[eq.~(8)]{Haenggi2009A}, the Laplace transform of the p.d.f. of the aggregate artificial noise~$I_A$ in~(\ref{eq:agg-int-an-st}) is given by
\begin{align}
\mathcal{L}_{I_A}\left(s\right)=\exp\left(-\frac{\lambda_lC_{\alpha,2}G_N^\frac{2}{\alpha}}{N}\left(N-1\right)^{1-\frac{2}{\alpha}}P_A^{\frac{2}{\alpha}}s^{\frac{2}{\alpha}}\right)\label{eq:lap-agg-int-an-st}
\end{align}
where~$C_{\alpha,2}$ is defined in~(\ref{eq:const-c-alpha-n}).

By applying the probability generating functional of a PPP~(see Definition A.5 in~\cite{Haenggi2009B}) to the inner expectation of~(\ref{eq:pso-as-ex}), we have
\begin{align}
p_{\rm so}&\!=\!1\!-\!\mathbb{E}_{\Phi_A}\!\left\{\!\exp\!\left(\!-\frac{\lambda_{e}}{N}\int_{\mathbb{R}^{2}}\Pr\!\left({\rm SIR}_{z}\!>\!\beta_{e}|\Phi_A\right)\!{\rm d}z\!\right)\!\right\}\!\;.
\end{align}
Invoking the bounding technique used in~\cite{Ganti2007,Zhou2011B}~(i.e., applying Jensen's inequality), we get the following upper bound:
\begin{align}
p_{\rm so}\!\leq\! p_{\rm so}^{\rm UB}&\!:=\!1\!-\!\exp\left(-\frac{\lambda_{e}}{N}\int_{\mathbb{R}^{2}}\Pr\left({\rm SIR}_{z}>\beta_{e}\right){\rm d}z\right)\nonumber\\
&\!=\!1\!-\!\exp\left(-\frac{\lambda_{e}}{N}\int_{\mathbb{R}^{2}}\mathcal{L}_{I_A}\left(\frac{\beta_{e}D_{oz}^{\alpha}}{G_NP_I}\right){\rm d}z\right)\;.
\end{align}
Plugging in~(\ref{eq:lap-agg-int-an-st}) and changing to a polar coordinate system to evaluate the integral yields the results in~(\ref{eq:pso-as-ub}).

By considering the nearest eavesdropper only, we now derive the secrecy outage probability lower bound in~(\ref{eq:pso-as-lb}). Denote the location of the nearest eavesdropper by~$n$. By~\cite[eq.~(2)]{Haenggi2005}, the distance between the typical transmitter and the nearest eavesdropper in~$\Phi_Z$, denoted by~$D_{on}$, is distributed according to the following p.d.f.:
\begin{align}
f_{D_{on}}\left(z\right)=\frac{2\pi\lambda_e}{N}z\,{\rm e}^{-\frac{\pi\lambda_e}{N}z^2}\;,~z>0\;.\label{eq:near-eve-dist-pdf-st}
\end{align}
The received SIR at this eavesdropper is given by
\begin{align}
{\rm SIR}_n={G_NP_IS_{on}D_{on}^{-\alpha}}{I_A^{-1}}\label{eq:sir-near-eve-st}
\end{align}
where~$S_{on}\sim{\rm Exp}\left(1\right)$,~$I_A$ is the aggregate artificial noise, defined in~(\ref{eq:agg-int-an-st}).

By~(\ref{eq:sir-near-eve-st}), the secrecy outage probability in~(\ref{eq:pso-as-ex}) can be lower bounded by
\begin{align}
p_{\rm so}\geq p_{\rm so}^{\rm LB}&:=\Pr\left({\rm SIR}_n>\beta_e\right)\nonumber\\
&=\mathbb{E}_{D_{on},I_A}\left[\exp\left(-\frac{\beta_eD_{on}^\alpha}{G_NP_I} I_A\right)\right]\nonumber\\
&=\mathbb{E}_{D_{on}}\left[\mathcal{L}_{I_A}\left(\frac{\beta_eD_{on}^\alpha}{G_NP_I}\right)\right]\;.
\end{align}
Using~(\ref{eq:lap-agg-int-an-st}) and~(\ref{eq:near-eve-dist-pdf-st}) to evaluate the last expectation yields the results in~(\ref{eq:pso-as-lb}).

\subsection{Proof of Lemma~\ref{thm:PCO-INT}}\label{prf:PCO-INT}
Here we derive the interference distribution in~(\ref{eq:pdf-pi-bf}). For a given realization of the intended channel~${\bf h}$ and with the beamforming strategy in~(\ref{eq:an-an}), the covariance matrix of the transmitted signal vector~${\bf x}$ is
\begin{align}
{\bf C}_{\bf x}=P_I\frac{{\bf h}{\bf h}^H}{\|{\bf h}\|^2}+\frac{P_A}{N-1}{\bf W}{\bf W}^H\;.
\end{align}
Noting that~${\bf W}{\bf W}^H={\bf I}-\frac{{\bf h}{\bf h}^{H}}{\|{\bf h}\|^2}$, we can rewrite the covariance matrix as
\begin{align}
{\bf C}_{\bf x}=\frac{P_A}{N-1}{\bf I}+\left(P_I-\frac{P_A}{N-1}\right)\frac{{\bf h}{\bf h}^{H}}{\|{\bf h}\|^2}\;.
\end{align}
If this signal~${\bf x}$ is received by a non-intended receiver through an unknown channel~${\bf h}_z$, the corresponding interference power is given by
\begin{align}
P_x=\frac{P_A}{N-1}\|{\bf h}_z\|^2+\left(P_I-\frac{P_A}{N-1}\right){\left|{\bf h}_z^H\frac{\bf h}{\|{\bf h}\|}\right|^2}\;.
\end{align}

For a given~${\bf h}\neq{\bf 0}$ and with a random~${\bf h}_z$, the interference power~$P_x$ can be interpreted as the weighted sum of the squared magnitude of~${\bf h}_z$ and the squared magnitude of the projection of~${\bf h}_z$ on the direction of~${\bf h}$. Hence, the amplitude of~${\bf h}$ is irrelevant and only its direction matters; more importantly, with the ergodicity of~${\bf h}_z$ over the symmetric complex space, the statistical distribution of~$P_x$ becomes independent of the direction of~${\bf h}$. For this reason, we set~${\bf h}=\left[1,0,\ldots,0\right]^T$ to characterize the distribution of~$P_x$. Denote the elements of~${\bf h}_z$ as~$h_{zi}$ with~$i=1,\ldots,N$. The interference power~$P_x$ can be expressed as
\begin{align}
P_x&=P_I\left|h_{z1}\right|^2+\frac{P_A}{N-1}\sum_{i=2}^N\left|h_{zi}\right|^2\;.\label{eq:pi-split-an}
\end{align}
We first consider the case where~$P_I\neq\frac{P_A}{N-1}$~(i.e.,~$\phi\neq\frac{1}{N}$). Note that~$P_I\left|h_{z1}\right|^2\sim{\rm Exp}\left(\frac{1}{P_I}\right)$ and~$\frac{P_A}{N-1}\sum_{i=2}^N\left|h_{zi}\right|^2\sim{\rm Gamma}\left(N-1,\frac{P_A}{N-1}\right)$ are mutually independent. The p.d.f. of~$P_x$ can be computed as
\begin{align}
f_{P_x}\left(z\right)&=\frac{1}{P_I}\left(1-\frac{P_A}{\left(N-1\right)P_I}\right)^{1-N}{\rm e}^{-\frac{z}{P_I}}\nonumber\\
&\times\gamma\left(N-1,\left(\frac{N-1}{P_A}-\frac{1}{P_I}\right)z\right)\;,~z>0
\end{align}
where~$\gamma\left(\cdot,\cdot\right)$ is the regularized lower incomplete gamma function. Plugging in the series representation of~$\gamma\left(\cdot,\cdot\right)$ yields the results in~(\ref{eq:pdf-pi-bf}). Note that the expression above does not hold for~$P_I=\frac{P_A}{N-1}$~(i.e.,~$\phi=\frac{1}{N}$). From~(\ref{eq:pi-split-an}), it is clear that when~$\phi=\frac{1}{N}$,~$P_x\sim{\rm Gamma}\left(N,P_I\right)$.

\subsection{Proof of Theorem~\ref{thm:PCO-AN}}\label{prf:PCO-AN}
Here we derive the connection outage probability in~(\ref{eq:pco-an-ext-exp}). By~\cite[eq.~(8)]{Haenggi2009A}, the Laplace transform of the p.d.f. of the aggregate interference~$I_{IA}$ in~(\ref{eq:agg-int-an-all}) is given by
\begin{align}
\mathcal{L}_{I_{IA}}\left(s\right)=\exp\left(-\lambda_l\pi\mathbb{E}\left[P_x^{\frac{2}{\alpha}}\right]\Gamma\left(1-\frac{2}{\alpha}\right)s^{\frac{2}{\alpha}}\right)\;.\label{eq:lap-tran-I-pco}
\end{align}
If~$P_I\neq\frac{P_A}{N-1}$~(i.e.,~$\phi\neq\frac{1}{N}$), by~(\ref{eq:pdf-pi-bf}),~$\mathbb{E}\left[P_x^{\frac{2}{\alpha}}\right]$ is given in~(\ref{eq:expect-px});
\begin{figure*}[!t]
\begin{align}
&\mathbb{E}\left[P_x^{\frac{2}{\alpha}}\right]\!=\!\frac{1}{P_I}\left(1\!-\!\frac{P_A}{\left(N\!-\!1\right)P_I}\right)^{1-N}\left(P_I^{1+\frac{2}{\alpha}}\Gamma\left(1\!+\!\frac{2}{\alpha}\right)\!-\!\left(\frac{P_A}{N\!-\!1}\right)^{1+\frac{2}{\alpha}}\sum_{k=0}^{N-2}\left(1\!-\!\frac{P_A}{\left(N\!-\!1\right)P_I}\right)^k\frac{\Gamma\left(k+1+\frac{2}{\alpha}\right)}{\Gamma\left(k+1\right)}\right)\label{eq:expect-px}
\end{align}
\normalsize \hrulefill
\end{figure*}
and if~$P_I=\frac{P_A}{N-1}$~(i.e.,~$\phi=\frac{1}{N}$), we have
\begin{align}
\mathbb{E}\left[P_x^{\frac{2}{\alpha}}\right]&=P_I^{\frac{2}{\alpha}}\frac{\Gamma\left(N+\frac{2}{\alpha}\right)}{\Gamma\left(N\right)}\;.
\end{align}

In~(\ref{eq:sir-pco-an}),~$\|{\bf h}_o\|^2\sim{\rm Gamma}\left(N,1\right)$ and its complementary cumulative distribution function takes the form in~\cite[eq.~(9)]{Hunter2008}, with~$\mathcal{N}=\left\{1\right\}$,~$\mathcal{K}=\left\{1,\ldots,N-1\right\}$ and~$a_{nk}=\frac{1}{k!}$. Hence, with the Laplace transformation obtained in~(\ref{eq:lap-tran-I-pco}), we can invoke~\cite[Theorem~1]{Hunter2008} to characterize the connection outage probability as follows:
\begin{align}
p_{\rm co}&=1-\sum_{p=0}^{N-1}\left[\frac{\left(-s\right)^p}{p!}\frac{{\rm d}^p}{{\rm d}s^p}\mathcal{L}_{I_{IA}}\left(s\right)\right]_{s=\frac{\beta_br^\alpha}{P_I}}\label{eq:pco-an-ext}\;.
\end{align}
By convention, the zero-th order derivative denotes the function itself. Plugging~(\ref{eq:lap-tran-I-pco}) into~(\ref{eq:pco-an-ext}), we have the results in~(\ref{eq:pco-an-ext-exp}).

\subsection{Proof of Corollary~\ref{thm:PCO-LOW-AN}}\label{prf:PCO-LOW-AN}
Here we derive a low outage approximation for the connection outage probability in~(\ref{eq:pco-an-ext-exp}). From~(\ref{eq:pco-an-ext-exp}) and~(\ref{eq:psi-phi-an}), it is clear that when~$\psi\left(\phi\right)\to0$, the connection outage probability~$p_{\rm co}\to0$. Note that the quantity~$\psi\left(\phi\right)$ couples many system parameters like~$r$ and~$\lambda_l$, and more importantly,~$\phi$ and~$N$. By making~$\psi\left(\phi\right)$ small, we are effectively studying all cases of system parameters that may lead to a small~$p_{\rm co}$, including but not limited to the asymptotic regions where~$r\to0$ or~$\lambda_l\to0$. We expand~(\ref{eq:pco-an-ext-exp}) around~$\psi\left(\phi\right)=0$ as follows:
\small
\begin{align}
p_{\rm co}&\!=\!{\beta_b^\frac{2}{\alpha}}\psi\!\left(\phi\right)\!-\!{\rm e}^{-{\beta_b^\frac{2}{\alpha}}\!\psi\left(\phi\right)}{\beta_b^\frac{2}{\alpha}}\psi\!\left(\phi\right)\frac{2}{\alpha}\sum_{p=1}^{N-1}\frac{1}{p!}\zeta\!\left(p,1\right)\!+\!\mathcal{O}\left(\psi\!\left(\phi\right)^2\right)\nonumber\\
&\!=\!{\beta_b^\frac{2}{\alpha}}\!\psi\left(\phi\right)\!-\!{\beta_b^\frac{2}{\alpha}}\!\psi\left(\phi\right)\!\frac{2}{\alpha}\!\sum_{p=1}^{N-1}\!\frac{1}{p!}\!\prod_{l=1}^{p-1}\left(l\!-\!\frac{2}{\alpha}\right)\!+\!\mathcal{O}\left(\psi\left(\phi\right)^2\right)
\end{align}
\normalsize
where~$\zeta\left(\cdot,\cdot\right)$ is defined in~(\ref{eq:zeta-an}).
Ignoring the high order terms gives the results in~(\ref{eq:pco-an-app}).

\subsection{Proof of Theorem~\ref{thm:PSO-AN}}\label{prf:PSO-AN}
We first derive the secrecy outage probability upper bound in~(\ref{eq:pso-an-ub}). By~\cite[eq.~(8)]{Haenggi2009A}, the Laplace transform of the p.d.f. of the aggregate artificial noise~$I_A$ in~(\ref{eq:agg-int-an}) is given by
\begin{align}
\mathcal{L}_{I_A}\left(s\right)=\exp\left(-\lambda_lC_{\alpha,N}\left(\frac{P_A}{N-1}\right)^{\frac{2}{\alpha}}s^{\frac{2}{\alpha}}\right)\label{eq:lap-agg-an-an}
\end{align}
where~$C_{\alpha,N}$ was defined in~(\ref{eq:const-c-alpha-n}).

Using the generating functional of a PPP~(see Definition~A.5 in~\cite{Haenggi2009B}), by~(\ref{eq:pso-gen-1}), we have
\begin{align}
p_{\rm so}&=1-\mathbb{E}_{\Phi_L}\left\{\exp\left[-\lambda_e\int_{\mathbb{R}^2}\Pr\left({\rm SIR}_z>\beta_e|\Phi_L\right){\rm d}z\right]\right\}\nonumber\\
&\leq1-\exp\left[-\lambda_e\int_{\mathbb{R}^2}\Pr\left({\rm SIR}_z>\beta_e\right){\rm d}z\right]\label{eq:pso-gen-2}
\end{align}
where the second line is obtained by applying Jensen's inequality. By~(\ref{eq:sirz-an}), the probability inside the integral can be evaluated as
\small
\begin{align}
&\Pr\left({\rm SIR}_z>\beta_e\right)\label{eq:pso-derive-tmp}\\
=&~\mathbb{E}_{\|{\bf g}_{oz}\|^2,I_A}\left\{\exp\left(-\frac{\beta_eD_{oz}^\alpha}{P_I}\left(\frac{P_A}{N-1}\|{\bf g}_{oz}\|^2D_{oz}^{-\alpha}+I_A\right)\right)\right\}\nonumber\\
=&\left(\beta_e\frac{\phi^{-1}-1}{N-1}+1\right)^{1-N}\mathcal{L}_{I_A}\left(\frac{\beta_eD_{oz}^\alpha}{P_I}\right)\nonumber\\
=&\!\left(\!\beta_e\frac{\phi^{-1}-1}{N-1}+1\!\right)^{1-N}\!\!\exp\!\left(\!-\beta_e^{\frac{2}{\alpha}}D_{oz}^2\lambda_lC_{\alpha,N}\left(\frac{\phi^{-1}-1}{N-1}\right)^{\frac{2}{\alpha}}\!\right)\!\;.\nonumber
\end{align}
\normalsize
Plugging~(\ref{eq:pso-derive-tmp}) into~(\ref{eq:pso-gen-2}), after changing to a polar coordinate system, evaluating the integral gives the results in~(\ref{eq:pso-an-ub}).

By considering the nearest eavesdropper only, we now derive the secrecy outage probability lower bound in~(\ref{eq:pso-an-lb}). Denote the location of the nearest eavesdropper by~$n$. By~\cite[eq.~(2)]{Haenggi2005}, the distance between the typical transmitter and the nearest eavesdropper in~$\Phi_E$, denoted by~$D_{on}$, is distributed according to the following p.d.f.:
\begin{align}
f_{D_{on}}\left(z\right)=2\pi\lambda_ez\,{\rm e}^{-\pi\lambda_ez^2}\;,~z>0\;.\label{eq:near-eve-dist-pdf-an}
\end{align}
The received SIR at this eavesdropper is given by
\begin{align}
{\rm SIR}_n=\frac{P_IS_{on}D_{on}^{-\alpha}}{\frac{P_A}{N-1}\|{\bf g}_{on}\|^2D_{on}^{-\alpha}+I_A}\label{eq:sirn-an}
\end{align}
where~$S_{on}\sim{\rm Exp}\left(1\right)$,~$\|{\bf g}_{on}\|^2\sim{\rm Gamma}\left(N-1,1\right)$ and~$I_A$ is the aggregate artificial noise, defined in~(\ref{eq:agg-int-an}).

Then, the secrecy outage probability in~(\ref{eq:pso-gen-1}) can be lower bounded by
\begin{align}
p_{\rm so}&\geq p_{\rm so}^{\rm LB}\nonumber\\
&:=\mathbb{E}_{D_{on},I_A,\|{\bf g}_{on}\|^2}\left[\Pr\left({\rm SIR}_n>\beta_e|{D_{on},I_A,\|{\bf g}_{on}\|^2}\right)\right]\nonumber\\
&=\left(\beta_e\frac{\phi^{-1}-1}{N-1}+1\right)^{1-N}\!\!\!\mathbb{E}_{D_{on},I_A}\left[\exp\left(-\frac{D_{on}^\alpha\beta_e}{P_I}I_A\right)\right]\nonumber\\
&=\left(\beta_e\frac{\phi^{-1}-1}{N-1}+1\right)^{1-N}\!\!\mathbb{E}_{D_{on}}\left[\mathcal{L}_{I_A}\left(\frac{D_{on}^\alpha\beta_e}{P_I}\right)\right]\;.
\end{align}
Plugging in~(\ref{eq:lap-agg-an-an}) and~(\ref{eq:near-eve-dist-pdf-an}) to evaluate the last expectation leads to the results in~(\ref{eq:pso-an-lb}).

\subsection{Proof of Corollary~\ref{thm:PHI-OPT-ST}}\label{prf:PHI-OPT-ST}
Here we show that the optimal power allocation ratio that maximizes the secrecy transmission capacity lower bound in~(\ref{eq:stc-lb-st}) is unique. Define the following quantities:
\begin{align}
x&:=\left(N-1\right)^{1-\frac{2}{\alpha}}\left(\phi^{-1}-1\right)^{\frac{2}{\alpha}}\nonumber\\
\delta&:=\frac{\alpha}{2}\;.
\end{align}
With the notation defined in~(\ref{eq:phi-opt-a4-st-notations}), the lower bound in~(\ref{eq:stc-lb-st}) can be expressed as
\begin{align}
C_{\rm Sector}^{\rm LB}&=(1-\sigma)\lambda_{l}\left[\log_{2}\left(\frac{1+\left(\frac{\varrho}{1+x}\right)^{\delta}}{1+\left(\frac{\varsigma}{x}\right)^{\delta}}\right)\right]^{+}\;.
\end{align}
We assume that~$\varrho>\varsigma$, such that a positive~$C_{\rm Sector}^{\rm LB}$ can be achieved by choosing~$x$ from $\left(\frac{\varsigma}{\varrho-\varsigma},\infty\right)$. More discussions on the feasibility of a positive $C_{\rm Sector}^{\rm LB}$ can be found in~\cite{Zhang2013B}. Then, by the monotonicity of the logarithm function, maximizing~$C_{\rm Sector}^{\rm LB}$ is equivalent to maximizing the following function:
\begin{align}
f\left(x\right)&=\left(1+\left(\frac{\varrho}{1+x}\right)^{\delta}\right)\left(1+\left(\frac{\varsigma}{x}\right)^{\delta}\right)^{-1}\;.
\end{align}
The derivative of $f\left(x\right)$ w.r.t. $x$ is given by
\begin{align}
\frac{{\rm d}}{{\rm d}x}f\left(x\right)=\frac{\varsigma^{\delta}+\frac{\varrho^{\delta}\varsigma^{\delta}}{\left(1+x\right)^{\delta+1}}-\varrho^{\delta}\left(\frac{x}{1+x}\right)^{\delta+1}}{\delta^{-1}x^{\delta+1}\left(1+\left(\frac{\varsigma}{x}\right)^{\delta}\right)^{2}}\label{eq:der-stc-phi-as}
\end{align}
where the denominator is always positive. When increasing~$x$ from~$\frac{\varsigma}{\varrho-\varsigma}$ to infinity, the first two terms in the numerator decreases from~$\varsigma^{\delta}+\varrho^{-1}\varsigma^{\delta}\left(\varrho-\varsigma\right)^{\delta+1}$ to~$\varsigma^{\delta}$ and the third term increases from~$\varrho^{-1}\varsigma^{\delta+1}$ to~$\varrho^{\delta}$ monotonically. Since~$\varrho>\varsigma$, the derivative of~$f\left(x\right)$ is first positive and then negative on~$x\in\left(\frac{\varsigma}{\varrho-\varsigma},\infty\right)$. When increasing~$\phi$ from zero to one, the value of~$x$ decreases from infinity to zero monotonically. This implies that the derivative of~$C_{\rm Sector}^{\rm LB}$ w.r.t.~$\phi$ is first positive and then negative with increasing~$\phi$, and thus the optimal value of~$\phi$ is unique.

\bibliographystyle{IEEEtran}
\bibliography{Cited}

\newpage

\begin{IEEEbiography}[{\includegraphics[width=1in,height=1.25in,clip,keepaspectratio]{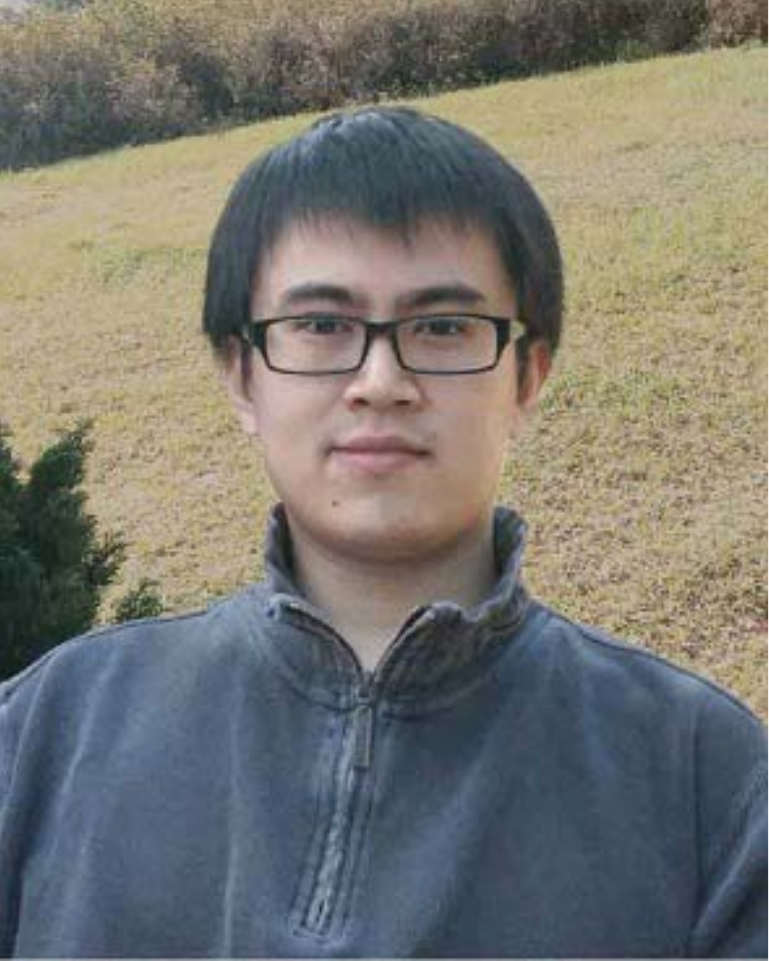}}]{Xi Zhang}~(S'11) received the B.E. degree in communication engineering from the University of Electronic Science and Technology of China in~2010. He is currently working toward the Ph.D. degree in electronic and computer engineering at the Hong Kong University of Science and Technology. His research interests are in the fields of wireless communication and signal processing techniques, including physical-layer security, ad hoc networking, and random matrix theory.
\end{IEEEbiography}

\begin{IEEEbiography}[{\includegraphics[width=1in,height=1.25in,clip,keepaspectratio]{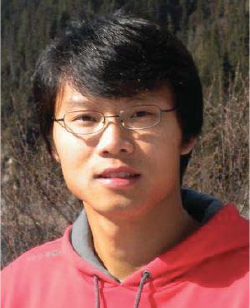}}]{Xiangyun Zhou}~(S'08-M'11) is a lecturer at the Australian National University~(ANU), Australia. He received the B.E.~(hons.) degree in electronics and telecommunications engineering and the Ph.D. degree in telecommunications engineering from the ANU in~2007 and~2010, respectively. From June~2010 to June~2011, he worked as a postdoctoral fellow at UNIK - University Graduate Center, University of Oslo, Norway. His research interests are in the fields of communication theory and wireless networks.

Dr. Zhou serves on the editorial board of the following journals: IEEE Communications Letters, Security and Communication Networks~(Wiley), and Ad Hoc \& Sensor Wireless Networks. He has also served as the TPC member of major IEEE conferences. Currently, he is the Chair of the ACT Chapter of the IEEE Communications Society and Signal Processing Society. He is a recipient of the Best Paper Award at the~2011 IEEE International Conference on Communications.
\end{IEEEbiography}

\begin{IEEEbiography}[{\includegraphics[width=1in,height=1.25in,clip,keepaspectratio]{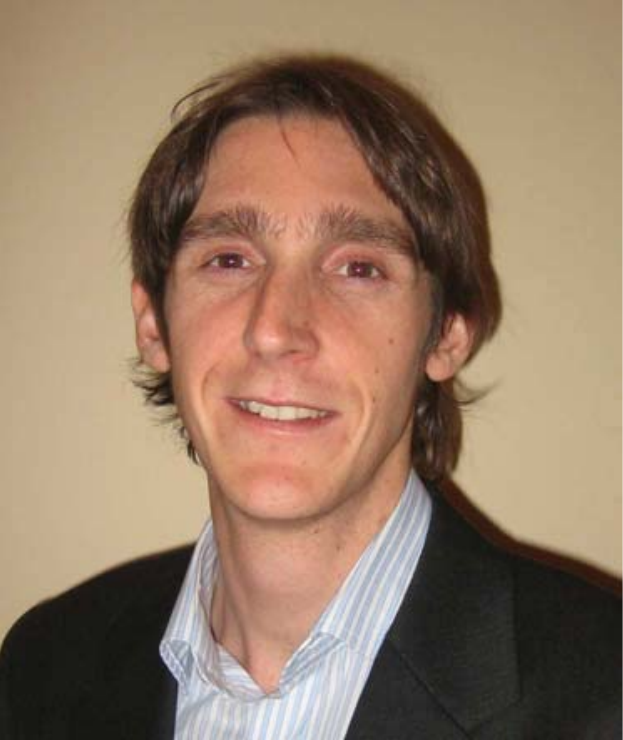}}]{Matthew R. McKay}~(S'03-M'07-SM'13) received the combined B.E. degree in electrical engineering and the B.IT. degree in computer science from the Queensland University of Technology, Australia, in~2002, and the Ph.D. degree in electrical engineering from the University of Sydney, Australia, in~2007. He then worked as a Research Scientist at the Commonwealth Science and Industrial Research Organization~(CSIRO), Sydney, prior to joining the faculty at the Hong Kong University of Science and Technology~(HKUST) in~2007, where he is currently the Hari Harilela Associate Professor of Electronic and Computer Engineering. He is also a member of the Center for Wireless Information Technology at HKUST, as well as an affiliated faculty member with the Division of Biomedical Engineering. His research interests include communications and signal processing; in particular the analysis and design of MIMO systems, random matrix theory, information theory, wireless ad-hoc and sensor networks, and physical-layer security.

Dr. McKay was awarded the University Medal upon graduating from the Queensland University of Technology. He and his coauthors have been awarded a Best Student Paper Award at IEEE ICASSP~2006, Best Student Paper Award at IEEE VTC~2006-Spring, Best Paper Award at ACM IWCMC~2010, Best Paper Award at IEEE Globecom~2010, Best Paper Award at IEEE ICC~2011, and was selected as a Finalist for the Best Student Paper Award at the Asilomar Conference on Signals, Systems, and Computers~2011. In addition, he received the~2010 Young Author Best Paper Award by the IEEE Signal Processing Society, the~2011 Stephen O. Rice Prize in the Field of Communication Theory by the IEEE Communication Society, and the~2011 Young Investigator Research Excellence Award by the School of Engineering at HKUST. Dr. McKay serves on the editorial boards of the IEEE Transactions on Wireless Communications and the mathematics journal, Random Matrices: Theory and Applications. In~2011, he served as the Chair of the Hong Kong Chapter of the IEEE Information Theory Society, whilst previously serving as the Vice-Chair and the Secretary. He has also served on the technical program committee for numerous international conferences, as well as the Publications Chair for IEEE SPAWC~2009, Publicity Chair for IEEE SPAWC~2012, and Poster Chair for IEEE CTW~2013.
\end{IEEEbiography}

\end{document}